\documentclass[aps,pra,reprint,showpacs,preprintnumbers,amsmath,amssymb,superscriptaddress]{revtex4-2}
\usepackage{eqnarray,amsmath}
\usepackage{graphicx}
\usepackage{amssymb}
\usepackage{textcomp}
\usepackage{color}
\usepackage{xcolor}
\usepackage{natbib}
\usepackage{dcolumn}
\usepackage{ragged2e}
\usepackage{array}
\usepackage{appendix}
\usepackage{etoolbox}

\begin{document}
\preprint{AP/Rb3+}

\title{Ultracold charged atom-dimer collisions: state-selective charge exchange and three-body recombination}

\author{A. Pandey}
\email{amrendra.pandey@universite-paris-saclay.fr } 
\affiliation{Universit\'e Paris-Saclay, CNRS, Laboratoire Aim$\acute{\text e}$ Cotton, 91405, Orsay Cedex, France}
\author{R. Vexiau}
\affiliation{Universit\'e Paris-Saclay, CNRS, Laboratoire Aim$\acute{\text e}$ Cotton, 91405, Orsay Cedex, France}
\author{L. G. Marcassa}
\affiliation{Instituto de F\'{i}sica de S\~{a}o Carlos, Universidade de S\~{a}o Paulo, Caixa Postal 369, 13560-970, S\~{a}o Carlos, S\~{a}o Paulo, Brazil}
\author{O. Dulieu}
\affiliation{Universit\'e Paris-Saclay, CNRS, Laboratoire Aim$\acute{\text e}$ Cotton, 91405, Orsay Cedex, France}
\author{N. Bouloufa-Maafa}
\affiliation{Universit\'e Paris-Saclay, CNRS, Laboratoire Aim$\acute{\text e}$ Cotton, 91405, Orsay Cedex, France}

\date{\today}

\begin{abstract} 
Based on an accurate determination of the potential energy surfaces of  Rb$_3^+$ correlated to its first asymptotic limit Rb$^+$$+$Rb($5s$)$+$Rb($5s$), we identify the presence of intersections of a pair of singlet and triplet surfaces over all interparticle distances, leading to Jahn-Teller couplings. We elaborate scenarios for charge exchange between ultracold charged atom-dimer complex (Rb$+$Rb$_2^+$ or Rb$^+$$+$Rb$_2$), predicting a strong selectivity on the preparation of the initial state of the dimer. We also demonstrate that the JT couplings must drive the three-body recombination (TBR) of Rb$^+$, Rb, and Rb at ultracold energies. Using the current analysis, we provide a consistent picture of the TBR experiments performed in ion-atom hybrid Rb samples \cite{dieterle2020inelastic,harter2012single}. We also demonstrate the presence of JT coupling as a general phenomenon in the singly-charged homonuclear alkali triatomic systems.
\end{abstract}
\maketitle

\section{Introduction}

The exquisite control of the preparation of ultracold atomic quantum degenerate gases in ongoing experiments opened the possibilities to drive the insertion of charged impurities, with the objective of observing various phenomena reflecting the sudden anisotropy generated by the charged particle \cite{tomza2019}. Up to now, due to the strong long-range interaction between the ionic and neutral particles varying as $R^{-4}$ (where $R$ is the interparticle distance), which dominates the van der Waals interaction (in $R^{-6}$) between the atoms, experiments involving hybrid ion-atom traps have observed the unavoidable process of three-body recombination (TBR), leading to the formation of a weakly-bound molecular ion further interacting with the atomic gas. In \cite{krukow2016energy,krukow2016reactive,mohammadi2021life}, a single cold $^{138}$Ba$^+$ ion is immersed in an ultracold Rb gas, leading to the formation of Rb$_2^+$ and RbBa$^+$ molecular ions. In \cite{dieterle2020inelastic,dieterle2021transport}, weakly-bound Rb$_2^+$ ions have been detected when a single cold $^{87}$Rb$^+$ ion is created inside a $^{87}$Rb Bose-Einstein condensate, thus invoking resonant charge-neutral interaction. In a previous experiment, the formation of deeply bound Rb$_2$ molecules has been indirectly observed by monitoring the decay of the atomic cloud after the production of highly energetic Rb$^+$ ions following TBR \cite{harter2012single,schmid2012apparatus,schmid2010}. The formation of deeply-bound Rb$_2$, however, is not compatible with the current understanding of ultracold TBR, which rather supports the formation of only weakly-bound dimers \cite{krukow2016energy,perez2021cold}. This apparent inconsistency may be due to the occurrence of a secondary process. It is worth noticing that such controlled hybrid systems may provide new insights on TBR, which is an essential process in many areas like the chemistry of the early universe \cite{dalgarno2005molecular,galli2013dawn,flower2007three,ramachandran2009revisiting,he2016global}, or cluster physics \cite{hauser2015classic,hauser2010jahn}.

In the present work, we identify charge exchange (CE) within a charged atom-dimer complex as the crucial process underlying the results of the experiments above, taking the example of the [Rb-Rb$_2$]$^+$ complex. We carried out accurate quantum chemistry calculations of the Rb$_3^+$ electronic states, revealing the presence of symmetry-required conical intersection where Jahn-Teller (JT) couplings act over all distances between the monomer and the dimer. Conical intersections have been observed in homonuclear alkali triatomic molecules, e.g., Li$_3$, Na$_3$, K$_3$, Rb$_3$ \cite{sadygov1999unusual, hauser2015classic, hauser2010jahn, schnabel2021towards}. Also, its role in the bond rearrangement reactions has been studied for many A$_2$B type alkali triatomic complexes \cite{fu2018accurate, he2016global, yin2020global, kendrick2021non, kendrick2021quantum, hermsmeier2021quantum, croft2017universality, jasik2018potential, hauser2011homo}. Studies involving the charged alkali trimer counterpart, A$_3^+$ and A$_2$B$^+$ \cite{ pavolini1987b,jeung1990lowest,smialkowski2020interactions}, along with studies on alkali clusters \cite{bonacic1991quantum,ray1989ab,ray1990ab,spiegelmann1988ab} are rather limited and primarily focus on the ground state properties. The occurrence of Jahn-Teller coupling in the charged alkali trimers has been noted in \cite{jeung1990lowest,pavolini1987b}. For Rb$_3^+$, along with other charged alkali trimers, a system with even number of valence electrons \cite{dillon2007seams, mozhayskiy2006conical, wu2019conical, zhu2016non}, non-adiabatic coupling terms involved in JT coupling remain strong in the absence of spin-orbit effects \cite{matsika2002spin}. This dynamical coupling is the primary direct mechanism that could induce CE within the complex. In some triatomic complexes, however, internal conversion or intersystem crossing due to accidental intersections could also be present \cite{yarkony1998conical}. In the present work, we predict that JT-induced CE is strongly selective with respect to the preparation of the quantum state of the dimer. We also argue that the JT coupling is active when the three ultracold particles Rb$^+$, Rb, and Rb are far from each other, thus being responsible for TBR.

In Section \ref{sec:computation}, we first present the geometries of  Rb$_3^+$ that are of particular interest for elaborating our approach. We briefly address the nature of the quantum chemistry calculations, which deliver the lowest potential energy surfaces (PESs) of Rb$_3^+$, that are examined for a series of representative geometries. In Section \ref{sec:jahnteller}, we characterize the electronic states that are effectively interacting via the JT coupling. In Section \ref{sec:ce}, we investigate the possible occurrences of CE in the [Rb-Rb$_2$]$^+$ complex, depending on the spatial geometry of the complex and the internal state of the dimer. In Section \ref{sec:disc}, we discuss the importance of these results to understand TBR, and we propose multi-step reactive paths to consistently interpret the experiments of  \cite{harter2012single,dieterle2020inelastic}. In Section \ref{sec:conclusion}, we summarize our results and elaborate on their possible future outcomes. In the rest of the paper, distances are expressed in units of Bohr radius $a_0$.

\section{Collision Geometries and Computational Methods}
\label{sec:computation}

\begin{figure}[t]
\begin{tabular}{cc}
\includegraphics[scale=0.5]{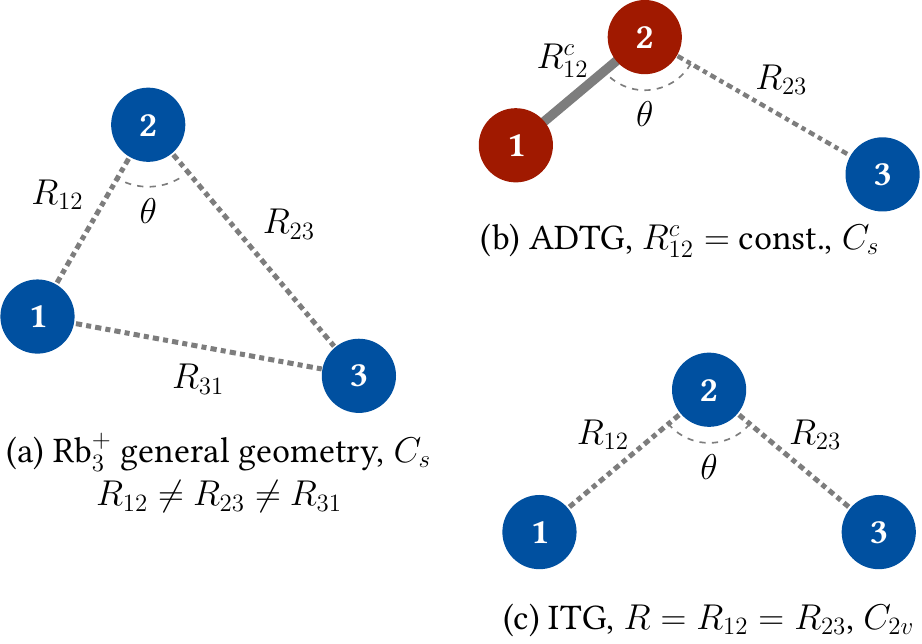}
\end{tabular}
\caption{Scheme of three Rb$^+$ cores marked as 1, 2, and 3. (a) The set of internal coordinates $R_{12}$, $R_{23}$, and $\theta$ ($\angle$123) chosen to represent the Rb$_3^+$ PESs. The third interparticle coordinate $R_{31}$ is indicated for further reference. (b) The atom-dimer triangle geometry (ADTG) belonging to the $C_{s}$ point group, with $R_{12}$ fixed to a given value $R_{12}^c$. (c) The isosceles triangle geometry (ITG) belonging to the $C_{2v}$ point group, with $R_{12} = R_{23} \equiv R$ for a given $\theta$.}
\label{fig:bmrng-geom}
\end{figure}

\begin{figure}[]
\begin{tabular}{cc}
\includegraphics[scale=0.3]{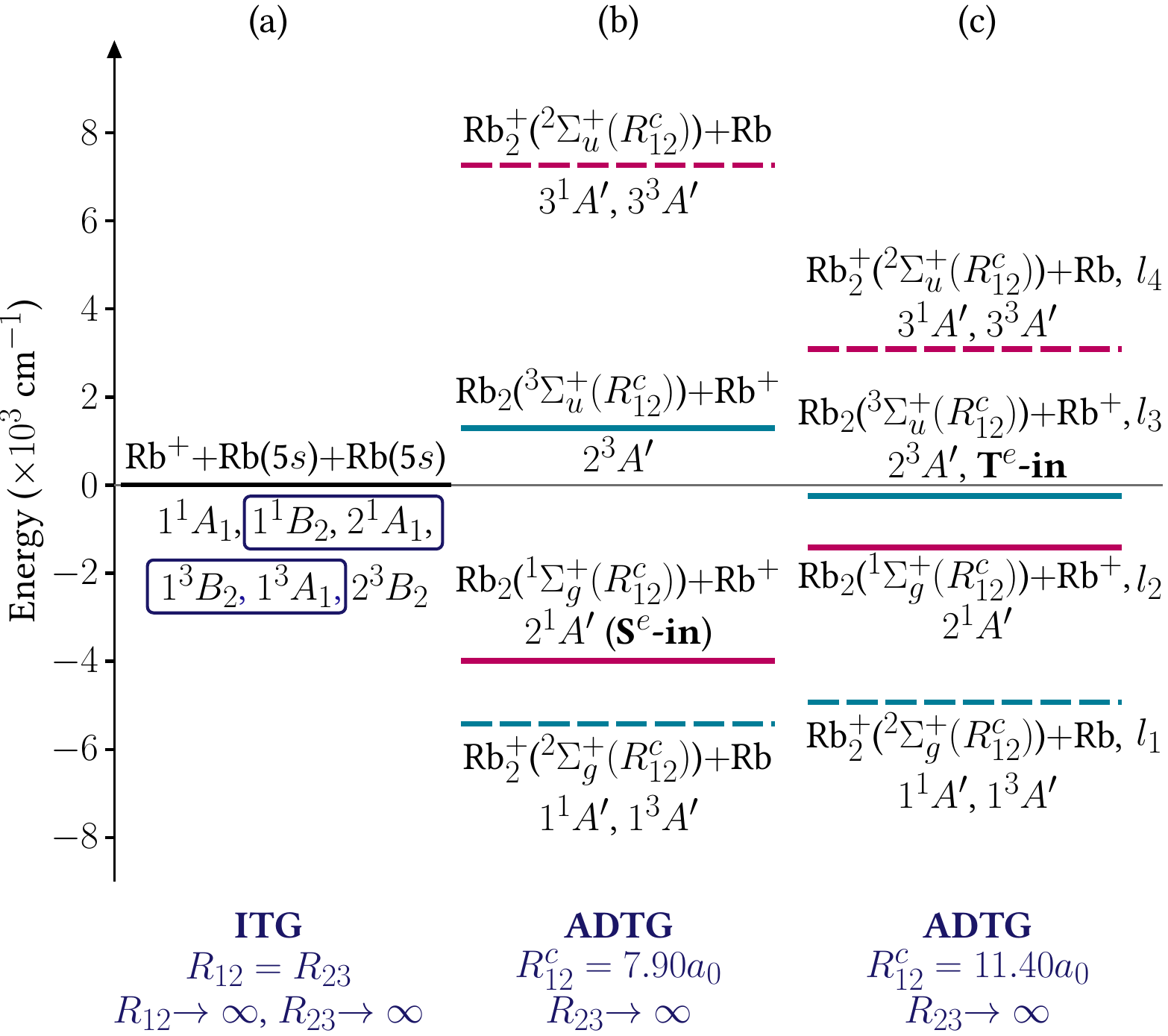}
\end{tabular}
\caption{Computed energies of Rb$_3^+$ dissociation limits ($R_{23}\rightarrow \infty$), with respect to the atomization energy of Rb$^+$+Rb($5s$)+Rb($5s$). (a) The six ITG (or $C_{2v}$) electronic states correlated to Rb$^+$+Rb($5s$)+Rb($5s$). Each rectangle specifies the pair of states that remain degenerate at $\theta=60$\textdegree~for finite $R$ values and are thus concerned with Jahn-Teller coupling. (b) Energies of the four associated monomer-dimer asymptotes (labeled as $l_1$, $l_2$, $l_3$, $l_4$), with the related ADTG (or $C_s$) states assuming $R^c_{12}$ fixed at the equilibrium distance ($7.90a_0$) of the $X^1\Sigma_g^+$ of Rb$_2$. (c) Same as (b) when $R^c_{12}$ is fixed at the equilibrium distance ($11.40a_0$) of the $a^3\Sigma_u^+$ state of Rb$_2$. The corresponding limits, identified as \textbf{S$^e$-in} and \textbf{T$^e$-in} respectively, figure possible initial geometries of the entrance channels.}
\label{fig:Rb3pasym}
\end{figure}

The Rb$_3^+$ PESs are expressed as functions of the three internal coordinates $R_{12}$, $R_{23}$, and $\theta$ (Fig. \ref{fig:bmrng-geom}). In the following, we will constantly refer to two specific geometries:
\begin{itemize}
    \item The atom-dimer triangle geometry (ADTG) with $R_{12}$ kept constant, of $C_s$ symmetry: this geometry is convenient to describe CE between the monomer and the dimer (both being either charged or neutral), so that $R_{12}$ is typically equal to the representative extension $R_{12}^c$ of a bound level of suitable molecular electronic state of the dimer, thus reflecting its initial preparation.
    \item The isosceles triangle geometry (ITG) with $R_{12} = R_{23} (\equiv R)$ for a given $\theta$ value, of $C_{2v}$ symmetry: at large distances, it will be helpful to investigate TBR. When $\theta=60$\textdegree, all $R_{ij}$ are equal (thus with $D_{3h}$ symmetry), that may lead to state degeneracies that are absent for all other geometries.
\end{itemize}

We carry on electronic structure calculations employing Multireference Configuration Interaction with single and double excitations (MRCI-SD) using the MOLPRO package \cite{werner1988efficient,MOLPRO-WIREs}. This approach is suitable to determine the electronic ground state and several excited states with consistent accuracy over all states. Each Rb$^+$ core is represented by the large effective core potential denoted as ECP36SDF accounting for all core electrons, with the recommended ($2s$, $2p$) Gaussian basis set \cite{szentpaly1982,fuentealba1983}. The associated core polarization potential (CPP) recommended in MOLPRO is added to model the core-valence electronic correlation, with parameters reported in the Appendix \ref{ssec:basis}. Thus the valence electrons of Rb$_3^+$ are explicitly treated, considering the above basis set augmented with a set of $1s$,$1p$, $2d$, $1f$ diffused functions that are optimized to reproduce Rb(5$^2$S$_{1/2}$) ionization potential and spectroscopic constants of Rb$_2$ $^1\Sigma_g^+$ ground state (See details in Appendix \ref{ssec:basis}). For Rb, Rb$_2$, Rb$_2^+$ electronic states, our calculations are extensively compared to the most recent ones, \cite{schnabel2022high}, and will appear in a separate paper \cite{dasilva2024}. In Appendix \ref{ssec:basis}, we argue that the present results are in good agreement with those of \cite{smialkowski2020} for the Rb$_3^+$ electronic ground state, showing its minima for ITG at $\theta = 60$\textdegree.

\begin{figure}[]
\begin{tabular}{c}
\includegraphics[scale=0.41]{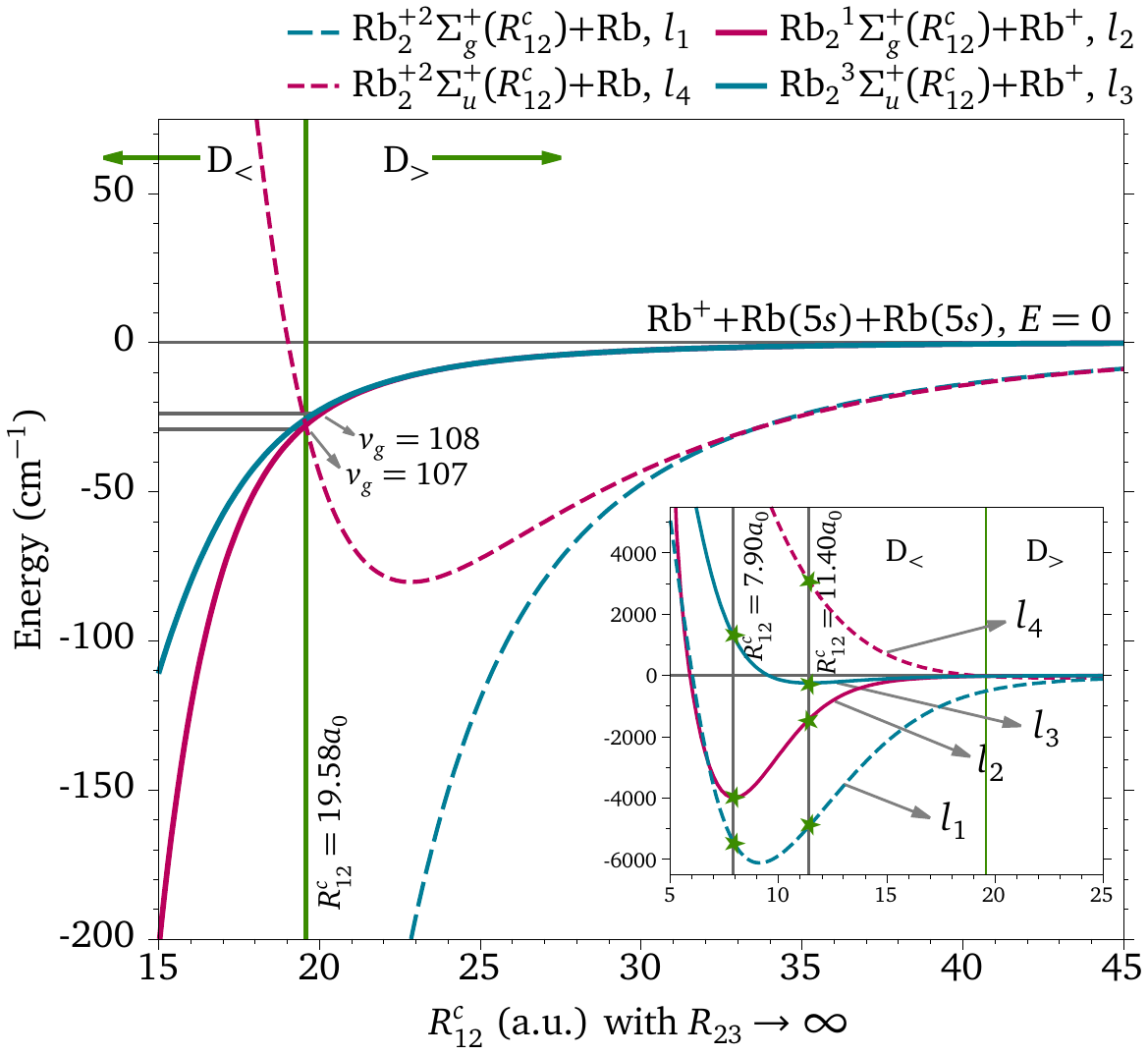}
\end{tabular}
\caption{Energies of the four Rb$_3^+$ ADTG asymptotes $l_1$, $l_2$, $l_3$, $l_4$, for a range of $R^c_{12}>15a_0$ values, with $R_{23}$$\rightarrow \infty$ (thus independent of $\theta$). The inset displays the entire graph, showing that these limits run along the PECs of the dimer quoted in the legend, with respect to the energy of the atomization limit Rb$^+$+Rb($5s$)$+$Rb($5s$). A vertical line at $R^c_{12}=19.58 a_0$, where $l_4$ crosses $l_2$ and $l_3$, defines the regions $D_>$ and $D_<$. For illustration purposes, the calculated $^{85}$Rb$_2(X^1\Sigma_g^+)$ vibrational levels in this vicinity are drawn. In the inset, the vertical lines at $R^c_{12}=7.90 a_0$ and $R^c_{12}=11.40 a_0$ cross the dimer PECs (stars) at energies of the ADTG asymptotes of Fig. \ref{fig:Rb3pasym} (b) and (c).}
\label{fig:Rb2np}
\end{figure}

In Fig. \ref{fig:Rb3pasym}, the computed energies of various dissociation limits of the lowest Rb$_3^+$ electronic states, relevant for the CE study, are displayed with respect to their atomization energy limit Rb$^+$+Rb($5s$)$+$Rb($5s$) taken as the zero of energy. Three singlet ($1^1A_1$, $1^1B_2$ $2^1A_1$) and three triplet ($1^3B_2$, $1^3A_1$, $2^3B_2$) states for ITG are correlated to this limit. Considering $R_{23}\rightarrow \infty$, two series of monomer-dimer dissociation limits are reported for ADTG. They correspond to $R_{12}^c = 7.90a_0$ and $R^c_{12}=11.40 a_0$, namely the equilibrium distance of the Rb$_2$ ground state $X^1\Sigma_g^+$ and lowest triplet state $a^3\Sigma_u^+$. They figure possible initial geometries respectively labeled as \textbf{S}$^e$-in and \textbf{T}$^e$-in (the index $e$ standing for dimer equilibrium distance) for the low-energy collision of Rb$^+$ with Rb$_2$ prepared in the lowest vibrational level of these states, involving the $2^1A'$ and $2^3A'$ Rb$_3^+$ electronic states. A singlet and a triplet $A'$ states are associated with both Rb$+$Rb$_2^+$ ADTG limits. On the other hand, Rb$^+$$+$Rb$_2$ ADTG asymptotes carry one $A'$ each. The four ADTG asymptotic limits for $R_{12}^c = 7.90a_0$ and $R^c_{12}=11.40 a_0$ are labeled in short and ordered as $l_1<l_2<l_3<l_4$. Note that $A''$ states are not relevant here as they are located at much higher energies, correlated to the second atomization energy limit Rb$^+$+Rb($5s$)$+$Rb($5p$).

For arbitrary fixed distances $R_{12}^c$ with $R_{23}\rightarrow \infty$, ADTG asymptotic limit energies run along the computed potential energy curves (PECs) of Rb$_2^+$($1^2\Sigma_g^+$), Rb$_2^+$($1^2\Sigma_u^+$), Rb$_2$($X^1\Sigma_g^+$), and Rb$_2$($a^3\Sigma_u^+$) (Fig. \ref{fig:Rb2np}). With respect to the atomization energy limit, the Rb$_2^+$($1^2\Sigma_u^+$) PEC crosses the $X^1\Sigma_g^+$ and $a^3\Sigma_u^+$ PECs of Rb$_2$ at $R^c_{12} = 19.58 a_0$, so that the ordering changes to $l_1<l_4<l_2<l_3$ beyond this distance. This will have important consequences for the CE dynamics. In the following we name $D_<$ and $D_>$ these two regions $R^c_{12} < 19.58 a_0$ and $R^c_{12} > 19.58 a_0$, respectively. In a classical analogue, $D_>$ ($D_<$) stands for Rb$_2$ prepared in high-lying (low-lying) vibrational levels.

In the next Sections, we will focus on representative geometries (ADTG and ITG) for given values of $\theta =$ 60\textdegree, 90\textdegree, 120\textdegree, 150\textdegree, 180\textdegree. Note that when $\theta <$60\textdegree, the internuclear distance $R_{31}$ becomes smaller than $R_{12}$ and $R_{23}$, so that the Rb$_3^+$ short-range PECs could be better described by $R_{31}$, $\theta$, and either $R_{12}$ or $R_{23}$.

\section{Jahn-Teller coupling in R\lowercase{b}$_3^+$ states}
\label{sec:jahnteller}

\begin{figure}[t]
\begin{tabular}{c}
\includegraphics[scale=0.28]{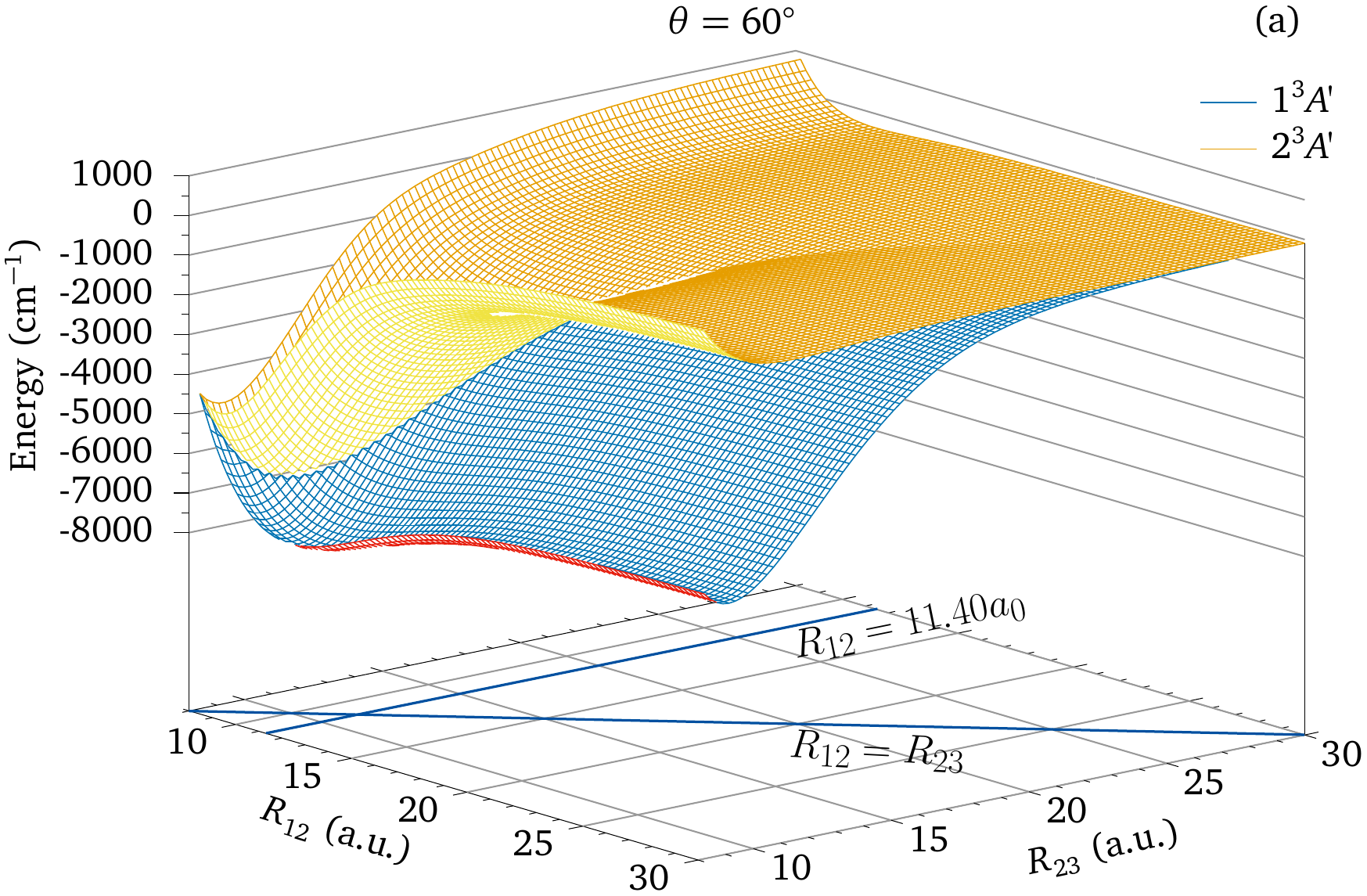} \\
\includegraphics[scale=0.45]{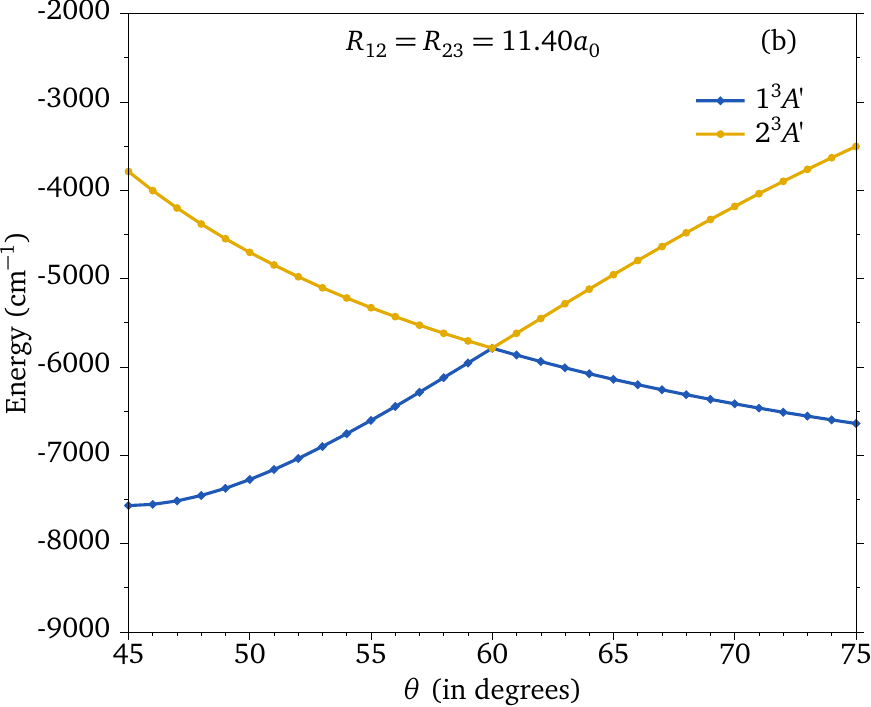}\tabularnewline
\end{tabular}
\caption{(a) Rb$_3^+$ PESs of the 1$^3A'$ and the 2$^3A'$ states as functions of $R_{12}$ and $R_{23}$, for $\theta=60$\textdegree. The states are degenerate along the diagonal line, $R_{12} = R_{23}$ (ITG), that defines the seam of the Jahn-Teller coupling. The other line in the horizontal plane shows ADTG for $R_{12} = 11.40 a_0$. (b) The same pair of states at $R_{12} = R_{23} = 11.40 a_0$ as a function of $\theta$.}
\label{fig:PESd1}
\end{figure}

\begin{figure*}[t]
\begin{tabular}{c}
\includegraphics[scale=0.270]{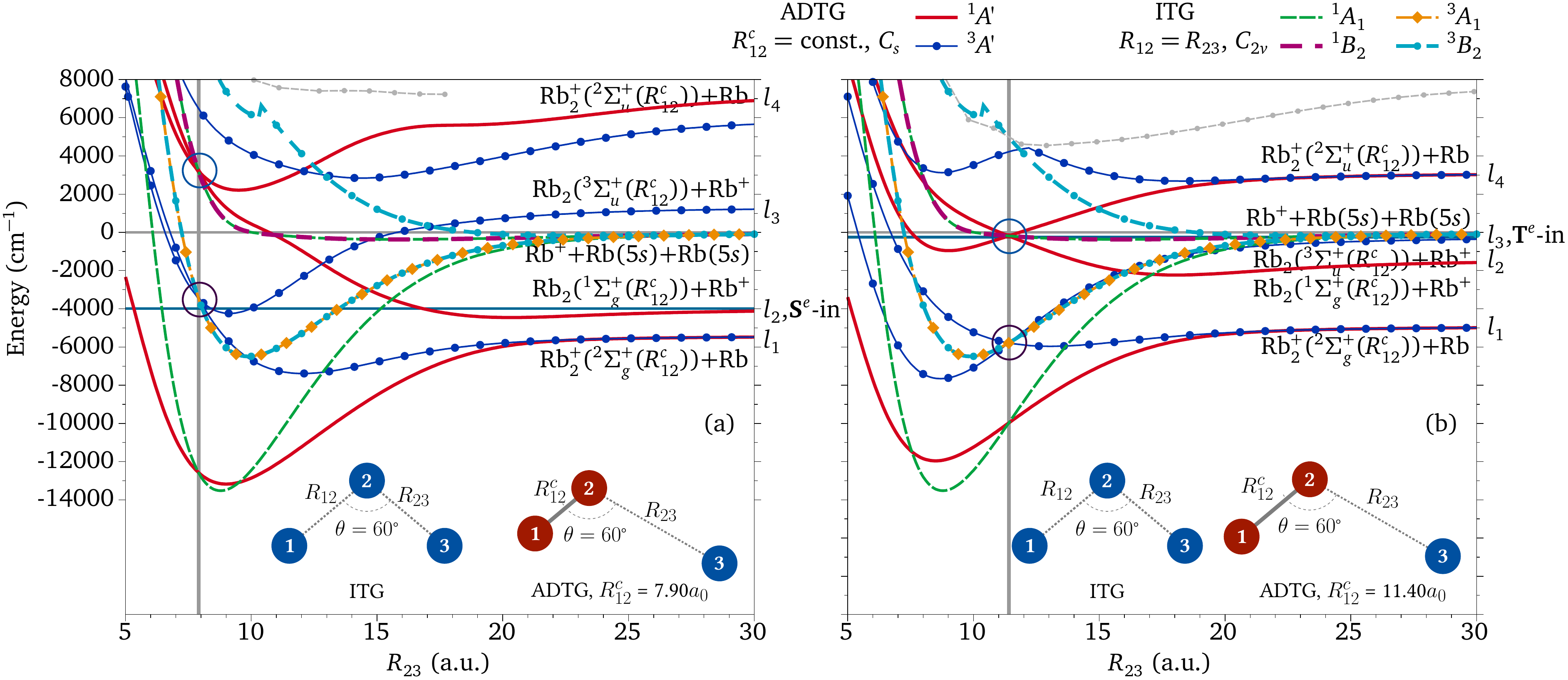}\tabularnewline
\end{tabular}
\caption{
One-dimensional (1D) PECs for ADTG ($C_s$ symmetry) at $\theta=60$\textdegree, (a) at the minimum $R_{12}^c = 7.90a_0$ of the $X^1\Sigma_g^+$ Rb$_2$ PEC, (b) at the minimum $R_{12}^c = 11.40a_0$ of the $a^3\Sigma_u^+$ Rb$_2$ PEC (vertical grey lines). Their dissociation limits are labeled as $l_1$, $l_2$, $l_3$, $l_4$ (see Fig. \ref{fig:Rb3pasym}). Occurrences of Jahn-Teller coupling are identified with circles. The corresponding curves for ITG ($C_{2v}$ symmetry) at $\theta=60$\textdegree~are drawn with dashed lines and are identical in both panels. A pair of ITG states (of $C_{2v}$ symmetry, which are degenerate at $\theta=60$\textdegree), namely $1^1B_2$, $2^1A_1$ (resp. $1^3B_2$, $1^3A_1$) have 1D curves which indeed crosses the pair of curves for ADTG states $2^1A'$, $3^1A'$ (resp. $1^3A'$, $2^3A'$) at the vertical lines, thus illustrating the seam of the Jahn-Teller coupling.  The energies of the possible entrance channels \textbf{S}$^e$-in and \textbf{T}$^e$-in (see Fig. \ref{fig:Rb3pasym}) for CE between Rb$^+$ and Rb$_2$, prepared in the lowest vibrational level of the  $X^1\Sigma_g^+$ and $a^3\Sigma_u^+$, are reported as horizontal indigo lines.}
\label{fig:PESb1}
\end{figure*}

As illustrated in Fig. \ref{fig:PESd1}, and Fig. \ref{fig:singlet_PES} in Appendix \ref{ssec:singletPES}, the geometries with $\theta=60$\textdegree~are the most remarkable. Two triplet ($1^3A'$,$2^3A'$) and two singlet ($2^1A'$,$3^1A'$) states are degenerate at all distances $R_{12}=R_{23}$, and match the pair of ITG states ($1^3B_2$, $1^3A_1$) and ($1^1B_2$, $2^1A_1$), respectively. Such a degeneracy reveals the occurrence of Jahn-Teller coupling. In many cases, JT-coupled pairs are correlated to a Rb$_2^+$+Rb and a Rb$^+$+Rb$_2$ asymptote, making the JT coupling the primary direct dynamical mechanism that can drive CE within the Rb$_3^+$ complex.

Having in mind the description of the CE process involving monomer-dimer dissociation limits, we explore in this Section one-dimensional cuts of the Rb$_3^+$ PESs to investigate in detail the JT coupling. Fig. \ref{fig:PESb1} displays such PECs for ADTG dissociating into a charged monomer-dimer complex, at $\theta = 60$\textdegree, and at the equilibrium distance of the $X^1\Sigma_g^+$ ($R_{12}^c = 7.90a_0$, panel (a)) and $a^3\Sigma_u^+$ ($R_{12}^c = 11.40a_0$, panel (b)) Rb$_2$ states. As expected, the $2^1A'$ and $3^1A'$ curves intersect each other at $R_{23} = R_{12}^c$, as well as the $1^3A'$ and $2^3A'$ curves, inducing JT coupling. PECs for the corresponding ITG (all dissociating into Rb$^+$+Rb(5s)+Rb(5s)), namely the ($1^1B_2$, $2^1A_1$) and the ($1^3B_2$, $1^3A_1$) pairs of PECs are displayed on the same plot for comparison purpose and are identical in both panels. These ITG states within each pair are degenerate for all values of $R_{23} = R_{12}$, thus representing the seam of the JT coupling in the singlet and triplet multiplicities.

\begin{figure*}[thb]
\begin{tabular}{c}
\includegraphics[scale=0.61]{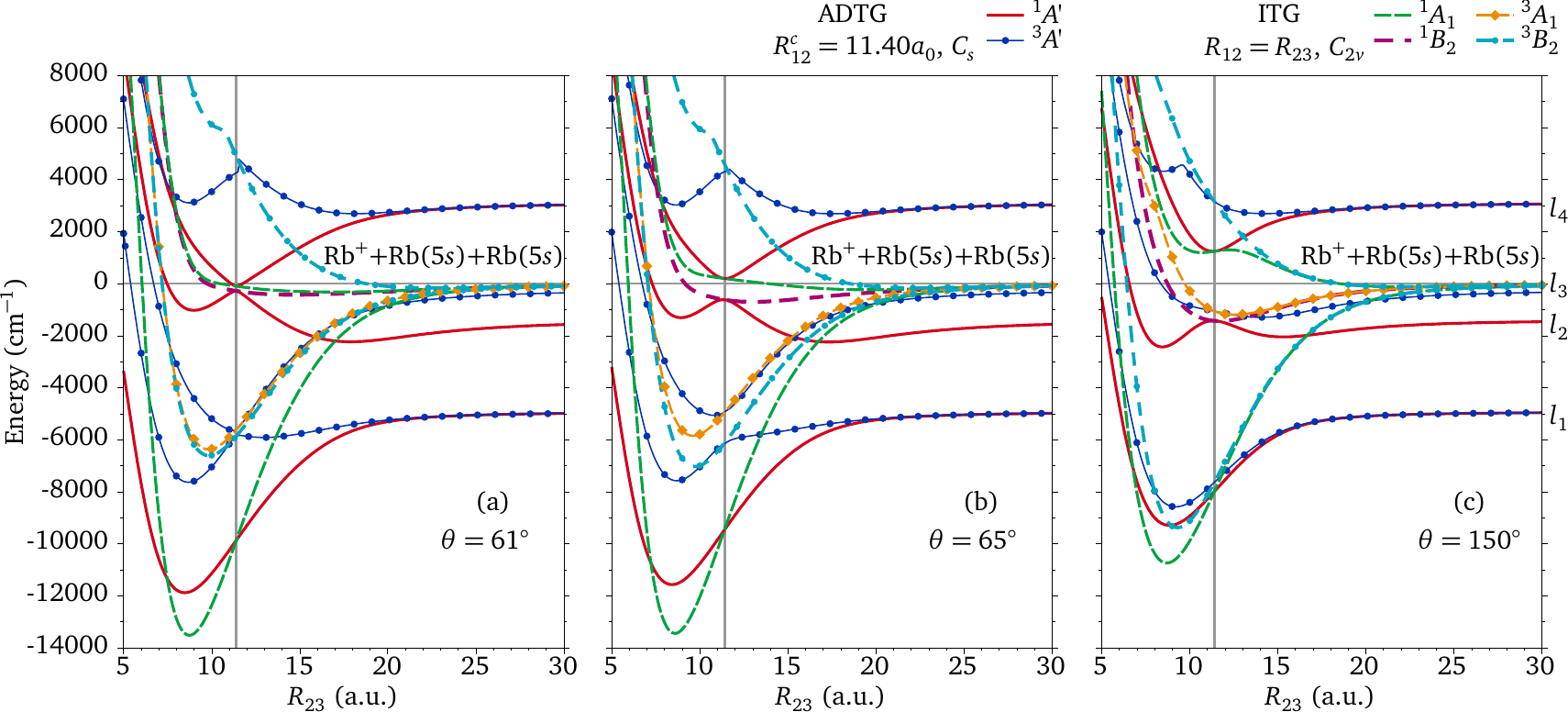}
\end{tabular}
\caption{One-dimensional cuts of the Rb$_3^+$ PESs (full lines) as functions of $R_{23}$, for ADTG ($C_s$ symmetry) at $R_{12}^c = 11.40a_0$ (vertical grey lines, at the minimum of the Rb$_2$, $a^3\Sigma_u^+$ PEC), (a) at $\theta=61$\textdegree, (b) $\theta=65$\textdegree, and (c) $\theta=150$\textdegree. The corresponding curves for ITG ($C_{2v}$ symmetry) are drawn with dashed lines for the same values of $\theta$.}
\label{fig:PESa1}
\end{figure*}

To complete Fig. \ref{fig:PESb1}, we draw in Fig. \ref{fig:PESa1} the same one-dimensional cuts of the Rb$_3^+$ PESs for several values of $\theta > 60$\textdegree, and at $R_{12}^c = 11.40a_0$. As expected, all circled curve crossings between ADTG curves of Fig. \ref{fig:PESb1} are significantly lifted as $\theta$ departs from 60\textdegree~by 1\textdegree~(see Fig. \ref{fig:PESa1}(a), also Fig. \ref{fig:PESd1}), as the Rb$_3^+$ complex has evolved in another direction of the configuration space. They now all appear as large avoided crossings. And, of course, the observed degeneracies of ITG pairs for the equilateral triangle geometry are also removed for $\theta \neq 60$\textdegree.

From the energy position of the \textbf{S}$^e$-in and \textbf{T}$^e$-in entrance channels in Fig. \ref{fig:PESb1}, we can already anticipate that when the Rb$_2$ molecule is prepared in the $X^1\Sigma_g^+$ lowest vibrational level, CE will be hardly possible, as the crossing between $^1A'$ curves at $\theta=60$\textdegree~lies at much higher energies than \textbf{S}$^e$-in. In contrast, its preparation in the $a^3\Sigma_u^+$ lowest vibrational level allows for CE, with the crossing between $^3A'$ located at lower energy than \textbf{T}$^e$-in: therefore, a strong selectivity of the CE process should be observable in these cases. However, the complex's ability to explore the minimal energy configuration also plays a crucial role. The following sections discuss CE scenarios for an arbitrary state preparation of Rb$_2$ (or Rb$_2^+$) dimer.

\section{State selectivity of charge exchange in the [R\lowercase{b}-R\lowercase{b}$_2$]$^+$ complex}
\label{sec:ce}

The occurrence of CE within the [R\lowercase{b}-R\lowercase{b}$_2$]$^+$ complex depends on two main features:
\begin{itemize}
    \item The change of the order of asymptotic limits $l_1$, $l_2$, $l_3$, $l_4$ (Fig. \ref{fig:Rb3pasym}), exemplified by the domains D$_<$ and D$_>$ in Fig. \ref{fig:Rb2np}.   
    \item The ability of the complex to explore the minimal energy configuration, given that JT coupling takes place at $\theta=60$\textdegree, starting from randomly aligned partners colliding at ultracold energies.
\end{itemize}

\begin{figure}[t]
\begin{tabular}{c}
\includegraphics[scale=0.335]{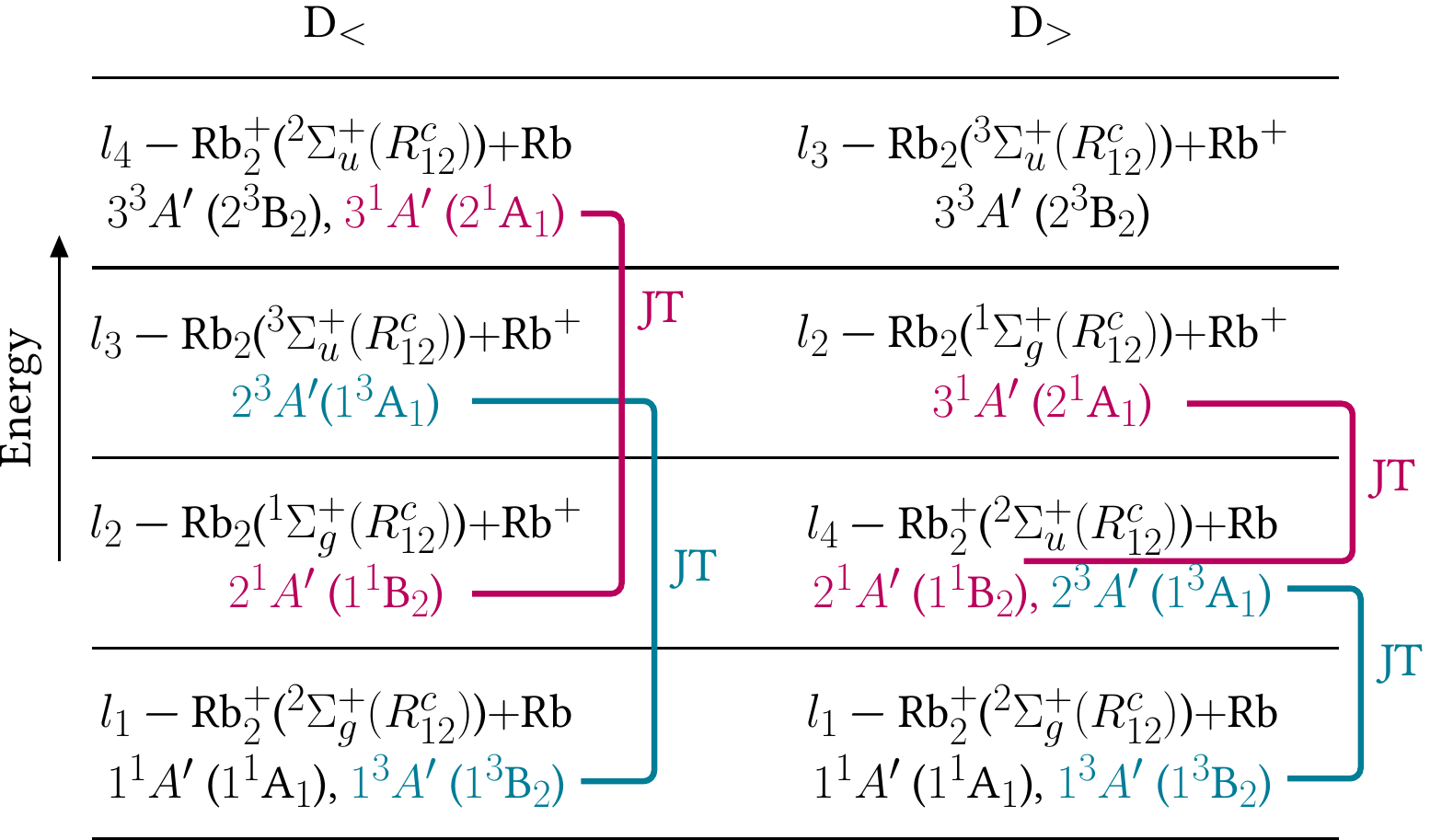}
\end{tabular}
\caption{The ordering in energy of the four monomer-dimer asymptotic limits $l_1$, $l_2$, $l_3$, $l_4$, depending on the distance domains D$_<$ and D$_>$ defined in Fig. \ref{fig:Rb3pasym}. The red and green brackets recall the ADTG (and the corresponding ITG) states coupled by the Jahn-Teller interaction.}
\label{fig:asydlim}
\end{figure}

First, CE possibilities are different in D$_<$ and D$_>$ regions due to the conditions arising from the two essential aspects of the Rb$_3^+$ states: (a) ADTG asymptotes correlated to Rb$_2^+$$+$Rb, $l_1$ and $l_4$, carry one singlet and one triplet Rb$_3^+$ states each; (b) JT coupling occurs between 2nd and 3rd singlet, ( $2^1A'$, $3^1A'$), and 1st and 2nd triplet, ($1^3A'$, $2^3A'$), Rb$_3^+$ states for all interparticle separations. These statements are summarized in Fig. \ref{fig:asydlim}, for convenience. Second, starting from a random orientation of the monomer-dimer at ultracold energy, the $\theta=60$\textdegree~should correspond to an energy minimum so that angular rearrangement during the collision drives the system toward the intersection of the PES relevant to the JT coupling.

\begin{figure}[t]
\begin{tabular}{c}
\includegraphics[scale=0.57]{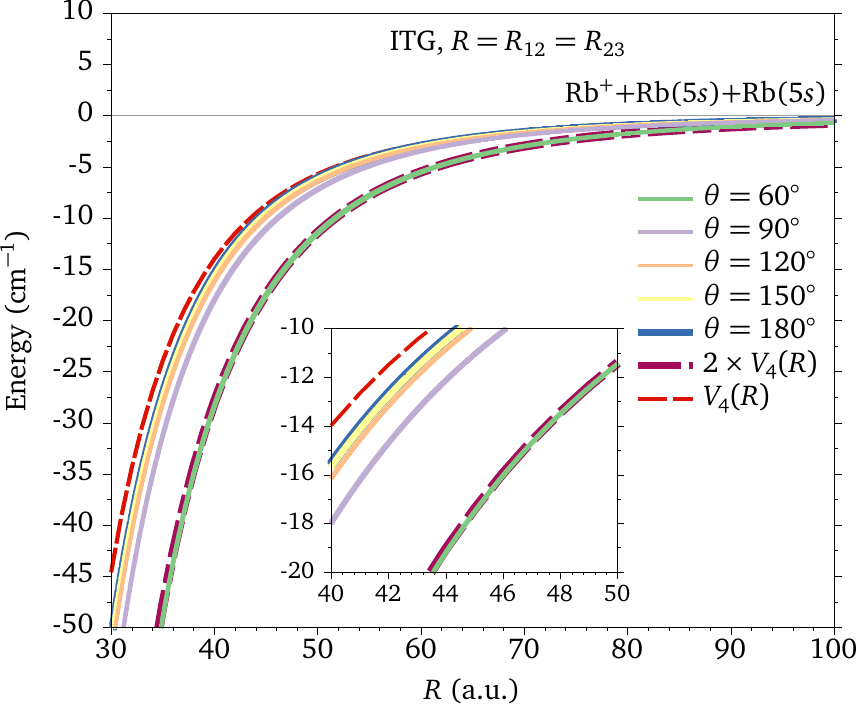}
\end{tabular}
\caption{One-dimensional cuts of the PESs of three Rb$_3^+$ ITG states ($1^1B_2$, $2^1A_1$, $1^3A_1$) involved in the JT coupling, for $R_{12}=R_{23} \equiv R \rightarrow \infty$, and for various values of $\theta$. The inset blows up a fraction of them for better visibility. The corresponding curves are degenerate at such distances and exhibit variations between $V_4(R)=-C_4/R^4$ and $2V_4(R)$.}
\label{fig:Rb3pc2vlr}
\end{figure}

\begin{figure}[h]
\begin{tabular}{c}
\includegraphics[scale=0.4]{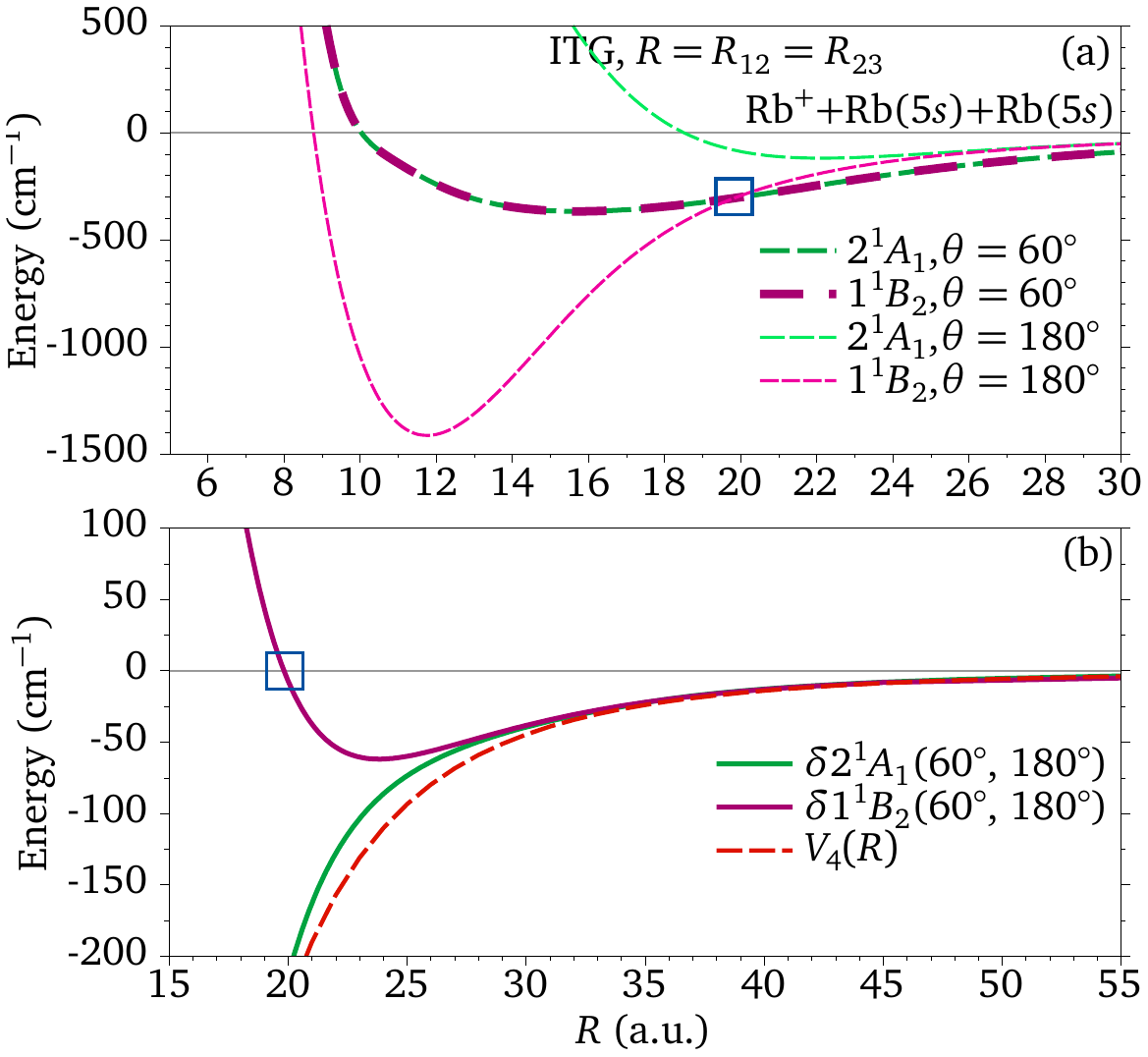}
\end{tabular}
\caption{(a) One-dimensional cuts of the PESs of the Rb$_3^+$ ITG states $1^1B_2$, $2^1A_1$ involved in the JT coupling, for $R_{12}=R_{23} \equiv R <30a_0$, and for $\theta=60$\textdegree~and 180\textdegree. Both curves are degenerate for 60\textdegree. Panel (b) shows the energy difference, $\delta$, between the PESs at $\theta=60$\textdegree~and 180\textdegree, highlighting that for the $1^1B_2$ curve, the latter geometry has a minimal energy for $R < 19.8 a_0$ (squared box). For the large $R$, $\delta$ for both states show $V_4(R)$ nature, as observed in Fig. \ref{fig:Rb3pc2vlr}. $2^1A_1$ shows its minima for 60\textdegree~geometry for all molecular sizes.}
\label{fig:Rb3c2vsmin}
\end{figure}

\begin{figure}[h]
\begin{tabular}{c}
\includegraphics[scale=0.4]{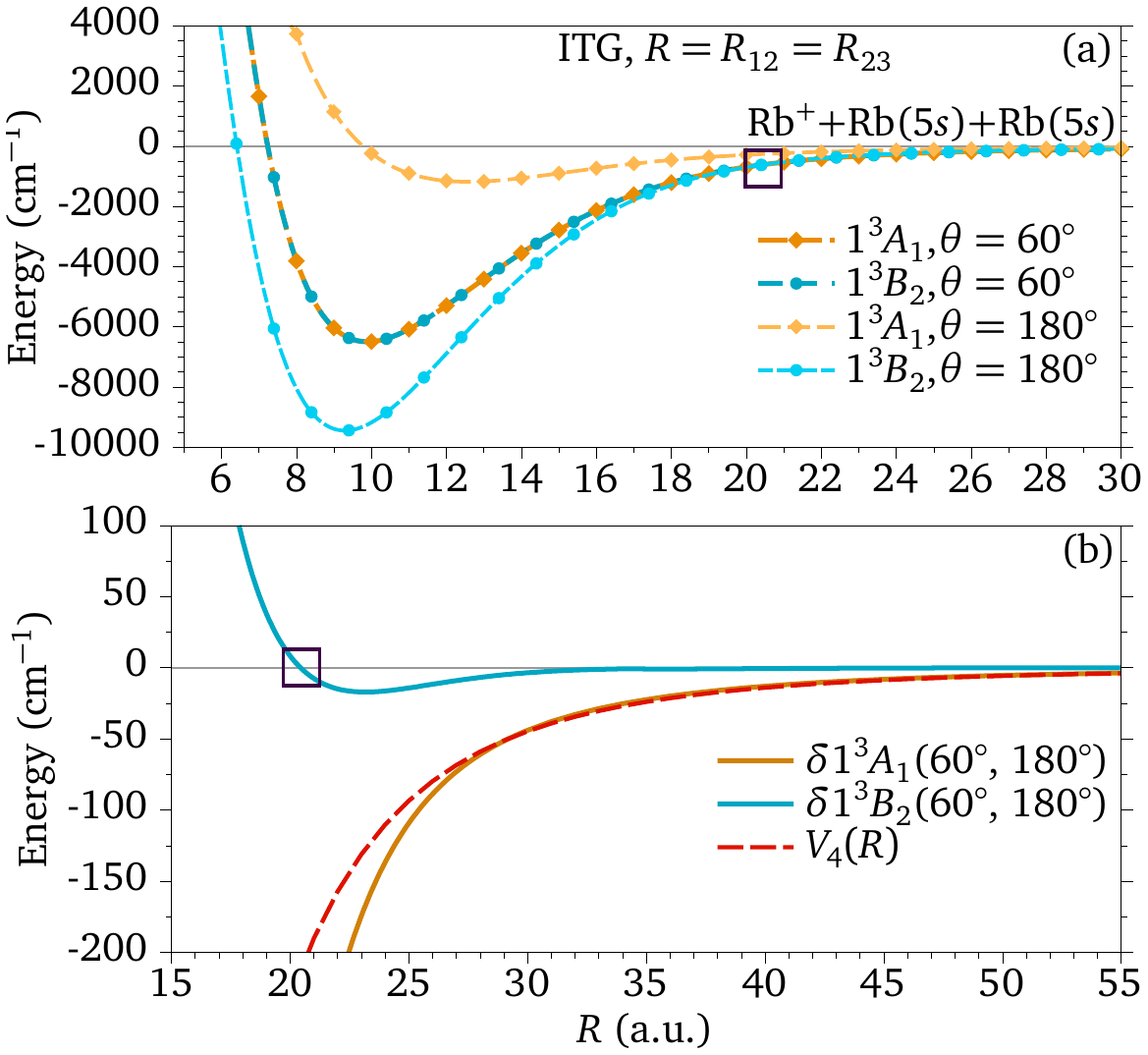}
\end{tabular}
\caption{(a) Same as Fig. \ref{fig:Rb3c2vsmin} for the $1^3B_2$ and $1^3A_1$ states. Both curves are degenerate for 60\textdegree. (b) For $R < 20.5 a_0$ (squared box), the $1^3B_2$ energy is minimal at 180\textdegree. For $R > 20.5 a_0$,  $\theta =60$\textdegree~becomes the minima for $1^3B_2$ with $\delta << V_4(R)$, referred as $\theta$-independent. $1^3A_1$ has its minima for $\theta =60$\textdegree~in both short and long range.}
\label{fig:Rb3c2vtmin}
\end{figure}

We first examine in Fig. \ref{fig:Rb3pc2vlr} the long-range behavior of the PESs of the ITG states for various angles $\theta \geq 60$\textdegree~ through one-dimensional cuts when $R_{12}=R_{23} \equiv R \rightarrow \infty$. The three JT curves of the $1^1B_2$, $2^1A_1$, and $1^3A_1$ ITG states are almost degenerate at such large distances and have their minimum for $\theta=60$\textdegree~ varying as $-2C_4/R^4$, with $2C_4 = \alpha_d=322.1$~a.u. is the dipole polarizability of Rb atom, derived from the present calculation, in good agreement with the experimental estimate (319.8~a.u., see \cite{gregoire2015measurements,maxwell2016analysis}). When $\theta$ increases, the curves are less attractive, approaching a variation as $-C_4/R^4$ at 180\textdegree. On the schematic picture of Fig. \ref{fig:bmrng-geom}, this implies that these states correspond to the Rb$^+$ ion as particle 1 (or 3). In contrast, the fourth ITG state $1^3B_2$ involved in JT coupling exhibits a long-range PEC varying as $-2C_4/R^4$ for any orientation: this suggests that particle 2 in the scheme of Fig. \ref{fig:bmrng-geom} is the Rb$^+$ ion. Overall, for such long-range approaches, the system will mostly evolve along the $\theta=60$\textdegree~geometry, thus favoring JT coupling.

If the dimer is prepared in a low vibrational state, it is likely that CE occurs at short monomer-dimer distances. Figures \ref{fig:Rb3c2vsmin} and \ref{fig:Rb3c2vtmin} display one-dimensional cuts of the singlet and triplet Rb$_3^+$ PESs for ITG (at $\theta=60$\textdegree~ and 180\textdegree). At such short distances, the upper states of the ITG coupled pairs, namely $2^1A_1$ and $1^3A_1$, have their minimal energy at $\theta=60$\textdegree~, thus favoring JT coupling toward the lower state of the pairs  $1^1B_2$ and $1^3B_2$. Thus, in the summary of Fig. \ref{fig:asydlim}, \ref{fig:Rb3c2vsmin}, and \ref{fig:Rb3c2vtmin}, at short distances, the ultracold CE is likely to occur starting from the $3^1A'$ and $2^3A'$ ADTG states, and not the reverse.

\begin{figure}[]
\begin{tabular}{c}
\includegraphics[scale=0.435]{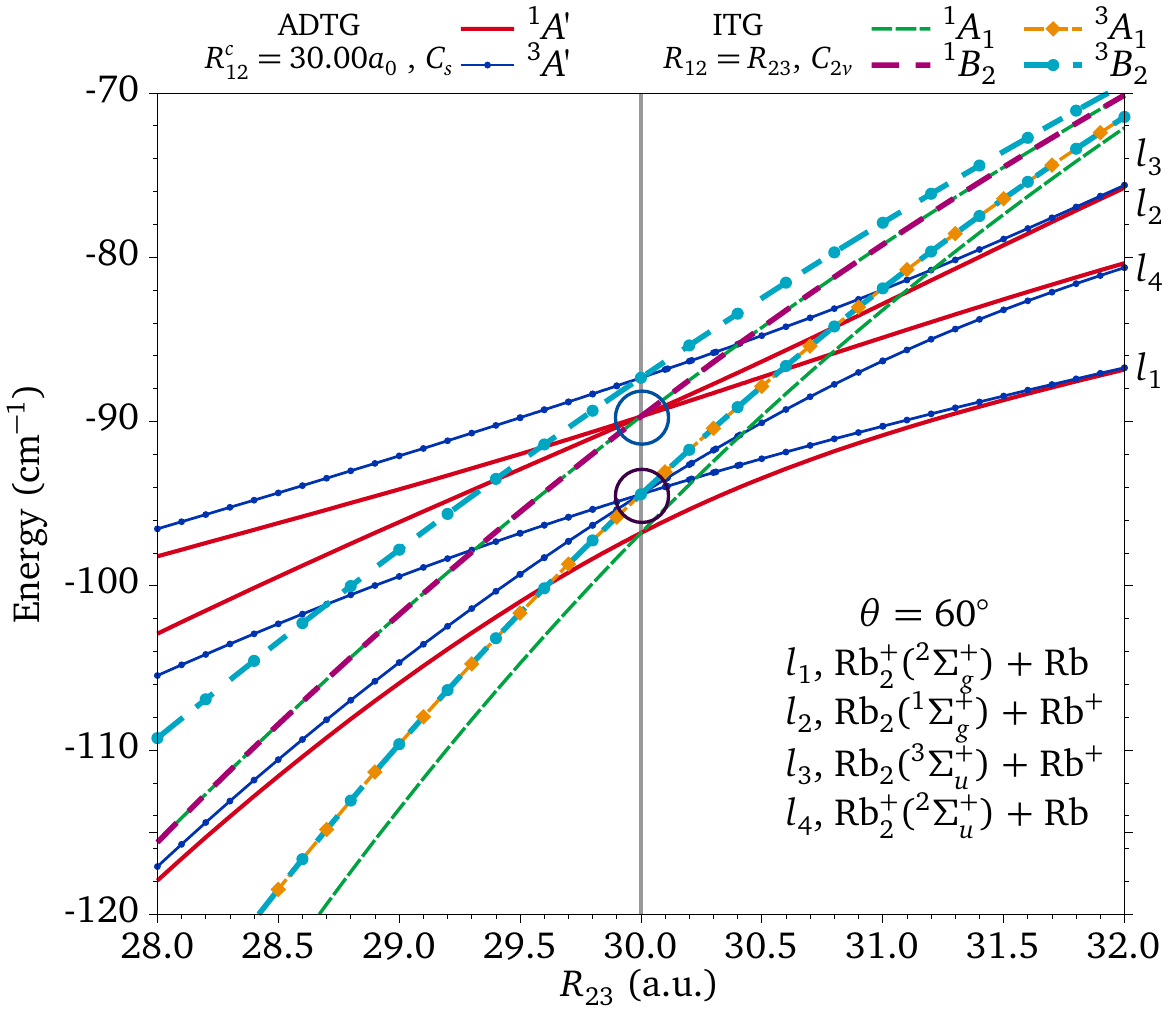}
\end{tabular}
\caption{At $\theta=60$\textdegree, ADTG for $R^c_{12}=30.00 a_0$ and ITG states as an example for the D$_>$ region. The order of the ADTG asymptotes is changed as compared to the cases of D$_<$ offered in Fig. \ref{fig:PESb1} (a) and (b). These curves are provided for $R_{23}$ range [5--30] $a_0$ in Fig. \ref{fig:lradtg30}, Appendix \ref{ssec:adtg30fr}. }
\label{fig:PESp1}
\end{figure}

Table \ref{tab:Rb3pgeo} summarizes all the possible scenarios for CE in the charged monomer-dimer complex involving the ADTG states, following the patterns discussed in the paragraphs above. A variety of cases is listed, first depending on the entrance channel characterized by one of the dissociation limits $l_1$, $l_2$, $l_3$, $l_4$ (Fig. \ref{fig:Rb3pasym}). The typical extension of the dimer, $R_{12}^c$, is given as belonging to the D$_<$ or D$_>$ domains (Fig. \ref{fig:Rb2np}), for which the exemplary cases of the ADTG PECs along with ITG states at $\theta=60$\textdegree~ are shown in Fig. \ref{fig:PESb1}(a) ,(b) (for $R^c_{12} = 7.90 a_0, 11.40 a_0$, $\in$D$_<$) and Fig. \ref{fig:PESp1} (for $R^c_{12} = 30.00 a_0$, $\in$D$_>$), respectively. Cases 1 and 3 represent the most striking result of this work: the strong selectivity of JT-induced CE at low energy, which is forbidden when the Rb$_2$ molecule is prepared in the lowest vibrational level or a low-lying one of its ground state $X^1\Sigma_g^+$, while being allowed for preparation in the $a^3\Sigma_u^+$ metastable state. Conversely, Rb$_2^+$ prepared in any level of its ground state $^2\Sigma_g^+$ is not exposed to CE (Cases 7 and 8), in contrast with a preparation in the $^2\Sigma_u^+$ state (Cases 5 and 6). In case 4, the asymptote Rb$_2$($a^3\Sigma_u^+$($R^c_{12}\in$D$_>$))$+$Rb$^+$, $l_3$, becomes the third triplet state of Rb$_3^+$, $3^3A'$, ( Fig. \ref{fig:asydlim}), which does not participate in the JT coupling, consequently unavailable to CE, also see Fig. \ref{fig:PESp1}. In case 8, on the other hand, JT coupling between states does not lead to the charge exchange but only to the electronic-vibrational energy transfer between Rb$_2^+$$^2\Sigma_g^+$ and $^2\Sigma_u^+$ states (Fig. \ref{fig:asydlim} and \ref{fig:PESp1}).

\begin{table*}[t]
\begin{ruledtabular}
\begin{tabular}{lc|lc|l}
Input collision channel & & Output collision channel&  &  CE  \\ 
&  &  & &\\ \hline
1. Rb$_2$($X^1\Sigma_g^+$($R^c_{12} \in$D$_<$)) $+$ Rb$^+$& $2^1A'$(180\textdegree~$\forall$ $R < 19.8 a_0$)& Rb$_2^+$ ($^2\Sigma_u^+$) $+$ Rb & $3^1A'$ & No \\
2. Rb$_2$($X^1\Sigma_g^+$($R^c_{12}\in$D$_>$)) $+$ Rb$^+$& $3^1A'$(60\textdegree~$\forall$ $R$)& Rb$_2^+$ ($^2\Sigma_u^+$) $+$ Rb &  $2^1A'$ & Yes \\
&  &  & &\\
3. Rb$_2$($a^3\Sigma_u^+$($R^c_{12} \in$D$_<$)) $+$ Rb$^+$& $2^3A'$(60\textdegree~$\forall$ $R$) & Rb$_2^+$ ($^2\Sigma_g^+$) $+$ Rb &  $1^3A'$ & Yes \\
4. Rb$_2$($a^3\Sigma_u^+$($R^c_{12}\in$D$_>$)) $+$ Rb$^+$& $3^3A'$ & - &  & No \\
&  &  & &\\  \hline
&  &  & &\\
5. Rb$_2^+$ ($^2\Sigma_u^+$($R^c_{12} \in$D$_<$)) $+$ Rb& $3^1A'$(60\textdegree~$\forall$ $R$)& Rb$_2$($X^1\Sigma_g^+$) $+$ Rb$^+$  &  $2^1A'$ & Yes \\
6. Rb$_2^+$ ($^2\Sigma_u^+$($R^c_{12}\in$D$_>$)) $+$ Rb&$2^1A'$(60\textdegree~$\forall$ $R > 19.8 a_0$) & Rb$_2$($X^1\Sigma_g^+$) $+$ Rb$^+$  & $3^1A'$ & Yes \\
&  &  & &\\
7. Rb$_2^+$ ($^2\Sigma_g^+$($R^c_{12} \in$D$_<$)) $+$ Rb&$1^3A'$(180\textdegree~$\forall$ $R < 20.5 a_0$) & Rb$_2$($a^3\Sigma_u^+$) $+$ Rb$^+$  &  $2^3A'$ & No \\
8. Rb$_2^+$ ($^2\Sigma_g^+$($R^c_{12}\in$D$_>$)) $+$ Rb&$1^3A'$($\theta$-independent~$\forall$ $R > 20.5 a_0$) & Rb$_2^+$($^2\Sigma_u^+$) $+$ Rb  & $2^3A'$ & No  \\
\end{tabular}
\caption{Overview of the 8 possible input dimer-monomer channels that could lead, or not, to charge exchange (CE), combining the information from Fig. \ref{fig:asydlim}, \ref{fig:Rb3pc2vlr}, \ref{fig:Rb3c2vsmin}, and \ref{fig:Rb3c2vtmin}. The assumed prepared internal state of the dimer is represented by its extension $R^c_{12} \in$D$_<$ or D$_>$. The involved ADTG electronic state with the position in $\theta$ of its minimum energy ITG configuration is displayed, recalling that CE will be possible via JT coupling only for $\theta= 60$\textdegree. In case 4 the ADTG state $3^3A'$ is not connected to any other state via JT coupling. In case 8, the $1^3A'$ state has no preferred geometry, while JT coupling between states does not lead to the charge exchange but only the electronic-vibrational energy transfer between Rb$_2^+$$^2\Sigma_g^+$ and $^2\Sigma_u^+$ states.}
\label{tab:Rb3pgeo}
\end{ruledtabular}
\end{table*}

\section{Perspectives for Three-body recombination of R\lowercase{b}$^+$$+$R\lowercase{b}$+$R\lowercase{b} in the ultracold regime}
\label{sec:disc}

\begin{figure*}[ht]
\begin{tabular}{c}
\includegraphics[scale=0.375]{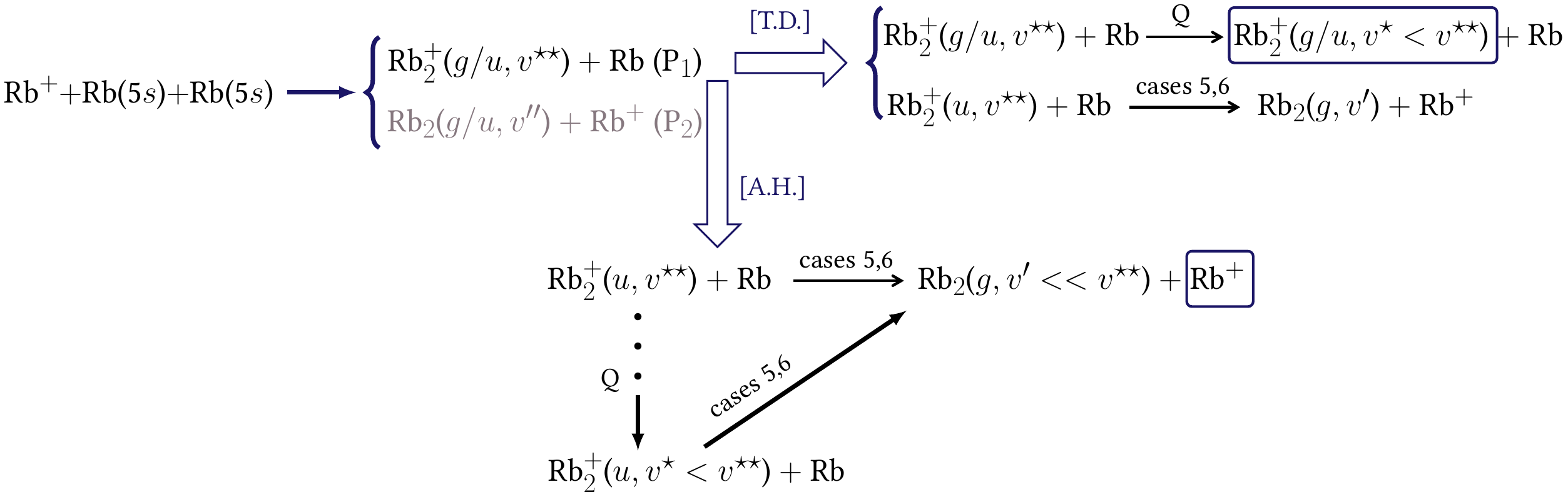}
\end{tabular}
\caption{Proposed reaction chain to interpret the two experiments reported in \cite{harter2012single,dieterle2020inelastic}. The $g$ and $u$ labels stand for the relevant $\Sigma_g^+$ and $\Sigma_u^+$ molecular states of the relevant molecules. The two possible TBR reaction outcomes are labeled with (P$_1$) and (P$_2$), the former being the most probable one (see text). The brackets [A.H.] and [T.D.] refer to \cite{harter2012single} and \cite{dieterle2020inelastic}, respectively. Q stands for vibrational quenching, and cases 5 and 6 refer to CE reactions from Table \ref{tab:Rb3pgeo}. The detected products are highlighted by rectangles.}
\label{fig:TBR-chain}
\end{figure*}

In the previous discussions, we discovered that in Rb$_3^+$, a pair of singlet states and a pair of triplet states have PESs that intersect each other within each pair for a particular geometry, namely ITG at $\theta=60$\textdegree, from short to large interparticle distances (Figs. \ref{fig:PESd1} and \ref{fig:singlet_PES}). Such a degeneracy thus induces JT coupling, which has important consequences for the dynamics of the monomer-dimer complex at low energies, which are summarized in Table \ref{tab:Rb3pgeo}. In particular, a strong selectivity is predicted for CE in [Rb$+$Rb$_2$]$^+$ collisions, depending on the preparation of the Rb$_2$,Rb$_2^+$ internal state.

As this pattern in the PESs is present at all interparticle distances, it is likely that TBR in Rb$^+$$+$Rb$+$Rb in the ultracold regime is also controlled by the resulting JT coupling. As three Rb$_3^+$ ITG states, $1^1B_2$, $2^1A_1$, and $1^3A_1$, have minimal energy for $\theta=60$\textdegree, where JT coupling takes place (the fourth PES, $1^3B_2$, being $\theta$-independent), it is likely that the approach at large distances of the three particles proceeds along this preferred geometry so that the JT coupling controls the dynamics at any distance. It is consistent with the prediction of the creation of a weakly-bound dimer during the TBR process \cite{harter2013population,krukow2016energy,perez2021cold}. To the best of our knowledge, there are only two experiments in the ultracold regime dealing with TBR in this particular ionic system Rb$_3^+$ \cite{harter2012single,dieterle2020inelastic}. Another one concerns the mixed complex Ba$^+$$+$Rb$+$Rb \cite{krukow2016energy,krukow2016reactive,mohammadi2021life}. The basic approach is to import a single trapped atomic ion inside an ultracold atomic gas. We briefly review them below, as they differ in their detection protocols, and we attempt to interpret them based on the present work.

In 2012, H\"arter \emph{et al.} \cite{harter2012single} reported on the observation of highly energetic Rb$^+$ ions (with kinetic energy between $h c \times 80$ and $h c \times 800$ cm$^{-1}$ ) produced after the introduction of a single cold ($\sim$ few mK) and trapped $^{87}$Rb$^+$ ion inside a quantum gas of $^{87}$Rb atoms at a temperature of $1.2 \mu$K. The time variation of the decay of the atomic cloud under the influence of a single trapped ion is consistent with the occurrence of a three-body collision and with the fact that the created ion is energetic enough to be expelled from the atomic gas (but not from the ionic trap), and then thermalized again down to its initial temperature after thousands of collisions with the atoms. The authors assessed that TBR should occur, creating deeply bound Rb$_2$ molecules, consistently with the high energy of the detected Rb$^+$ ions. They claim, however, that the formation of weakly-bound Rb$_2$ molecules could be possible but not detected, as the created Rb$^+$ ions would not leave the atomic gas and thus would not be detected. They also do not exclude the possibility of the formation of Rb$_2^+$.

Another experiment has been reported in 2020 by Dieterle \emph{et al.} \cite{dieterle2020inelastic}, with somewhat different initial conditions. A single cold and trapped $^{87}$Rb$^+$ ion with a kinetic energy $\sim k_B \times 50 \mu$K is immersed in a $^{87}$Rb Bose-Einstein condensate (BEC). The ion is created from a precursor Rb atom excited in a Rydberg state and field-ionized within the BEC, thus allowing for such ultracold energy. The authors measured that for each loss of a $^{87}$Rb$^+$, a molecular Rb$_2^+$ is observed. As predicted \cite{perez2021cold,perez2019vibrational}, they probe the creation of weakly-bound  Rb$_2^+$ ions immediately after the introduction of the ion, while over long observation time, a regime of secondary collisions between Rb$_2^+$ and Rb atoms is evidenced, leaving the molecular ion in relatively deeply-bound levels ($\sim -0.3$ cm$^{-1}$). Such a regime was also assessed in \cite{mohammadi2021life} for a Ba$^+$ ion immersed in a Rb BEC.

At first sight, the results of these two experimental works seem to disagree with each other, particularly for the formation of deeply bound Rb$_2$ and highly energetic Rb$^+$ ions in H\"arter \emph{et al.} \cite{harter2012single}. Our results are helpful in interpreting this apparent contradiction. 
First, \cite{perez2021cold} predicts only the production of weakly-bound dimers in the ultracold ion-atom-atom TBR processes. Second, due to the strength of the ion-atom interaction at long-range, larger than the atom-atom one, the formation of a molecular ion is favored. This is actually demonstrated in \cite{mirahmadi2023ion}, where for many ion-atom-atom cases, the TBR rate for weakly-bound molecular ion formation is found 4 orders of magnitude larger than the one for creating a weakly-bound neutral dimer. Starting from this hypothesis, the results of the two experiments above can be summarized according to the reaction chain displayed in Fig. \ref{fig:TBR-chain}. The TBR process first creates a weakly-bound Rb$_2^+$ either in its $^2\Sigma_g^+$ and $^2\Sigma_u^+$ state, labeled as (P$_1$). From Table \ref{tab:Rb3pgeo}, only the $^2\Sigma_u^+$ state thus leads to CE (cases 5, 6), possibly producing weakly-bound Rb$_2$ molecules, while Rb$_2^+$ ions in the $^2\Sigma_g^+$ state can be transferred to the $^2\Sigma_u^+$ state after a first Rb scattering via case 8, and then leading to CE. Formation of such weakly-bound Rb$_2$ molecules has not been directly reported in the experiment of \cite{harter2012single} as only hot Rb$^+$ ions are detected. It is also energetically allowed that the TBR generated Rb$_2^+$ molecular ion can undergo a series of Rb--Rb$_2^+$ quenching collisions before the CE collisions. Such intermediate quenching collisions for Rb$_2^+$ in $^2\Sigma_u^+$ state may not be effective in the production of highly energetic ions as its well depth is only 80.26 cm$^{-1}$ (Appendix \ref{ssec:basis}). On the other hand, Rb$_2^+$ ions in $^2\Sigma_g^+$, quenched deeply, are not open for the CE (case 7). However, these intermediate quenching collisions could well be responsible for the observation of Rb$_2^+$ with various internal energy in Dieterle \emph{et al.} \cite{dieterle2020inelastic}. Highly energetic Rb$^+$, observed in the H\"arter \emph{et al.} \cite{harter2012single}, could be arising from the events combining vibrational quenching and CE in Rb$_3^+$ complex. With the quenched Rb$_2^+$($^2\Sigma_u^+$) molecular ion colliding with Rb on the JT coupled singlet states, Rb$_3^+$ complex will have a substantial presence in the inner part of these PESs, for the interparticle separations $R$ upto $\sim 10 a_0$. Consequently, the CE collision events coupled with vibrational quenching could create Rb$_2$($^1\Sigma_g^+$) in the deeper vibrational levels and leave Rb$^+$ with large kinetic energies, which is observed in the H\"arter \emph{et al.} \cite{harter2012single}. Thus, both reported experiments can be interpreted within the framework proposed in this study.

TBR followed by the CE between atoms and molecular ions, summarized in Fig. \ref{fig:TBR-chain}, also solves another puzzle. It is noted that both experiments, \cite{harter2012single,dieterle2020inelastic}, produce similar TBR rate coefficients, in spite of the fact that in \cite{dieterle2020inelastic}, the Rb$_2^+$ signal is detected, while in \cite{harter2012single}, TBR is supposed to produce Rb$_2$ and Rb$^+$. As mentioned earlier, according to the simulations of \cite{mirahmadi2023ion}, the Rb$_2$$+$Rb$^+$ TBR outgoing channel should be $\sim 4$ orders of magnitude smaller than the TBR channel producing Rb$_2^+$$+$Rb. Therefore, the produced Rb$_2$ molecules in \cite{harter2012single} represent a strong case that state-selective charge exchange reactions discussed in the present manuscript are indeed very efficient. The rates of similar magnitude extracted in both experiments are then consistent: in both cases, the TBR signal is measured either by the direct detection of Rb$_2^+$ or by the indirect detection of the TBR products (neutral molecules).

\section{Conclusions and Outlook}
\label{sec:conclusion}

In this work, we have computed large portions of the potential energy surface manifold of the Rb$_3^+$ ion, correlated to the three-body break-up Rb($5s$)+Rb($5s$)+Rb$^+$. We discovered symmetry-required Jahn-Teller conical intersections between a pair of excited singlet and triplet states in Rb$_3^+$, occurring at equilateral triangle geometry where two PESs are degenerate. They dissociate into a charged atom-molecule pair, thus allowing for state-selective atom-dimer charge exchange at ultralow collision energies. This pattern is also found in Na$_3^+$ and K$_3^+$, and is probably present for all species. As the equilateral triangle geometry corresponds to a minimal energy for all interparticle distances in the long-range, we infer that this Jahn-Teller coupling is responsible for three-body recombination (TBR)  between two Rb($5s$) atoms and a Rb$^+$ ion into a Rb$_2^+$-Rb pair. The predictions arising from the JT-induced ion-atom-atom TBR and subsequent CE between atoms and molecular ions are helpful in finding a consistent picture of the charge-neutral dynamics at the ultracold temperatures, following the results of the two published experimental papers reporting on three-body recombination in this system \cite{harter2012single,dieterle2020inelastic}. We have evaluated the non-adiabatic couplings, which are found to be huge at the conical intersection, ensuring fast dynamics. Large cross-sections for charge exchange and TBR reactions are expected. In particular, the observed time-dependent Rb$_2^+$ yield shortly after the recombination in \cite{dieterle2020inelastic}, suggests that the TBR rate is at least as fast as the subsequent charge exchange between Rb$_2^+$ and Rb. The presence of JT-induced charge exchange channels could be investigated in hybrid traps dealing with homonuclear species like lithium \cite{niranjan2021}. Full quantum dynamical calculations could be envisioned, using, for instance, Multi-Conﬁguration Time-Dependent Hartree (MCTDH) method \cite{worth2000heidelberg}, and trajectory-based surface hopping techniques  \cite{mai2018nonadiabatic}. However, these calculations are usually involved and are beyond the scope of the present paper. On the other hand, several model systems with conical intersections provide sufficient evidence for the large cross-sections of the conical intersection-induced reactions. For instance, \cite{ferretti1997quantum} reports large transition probabilities  (up to 90-95\%) for two-state A-B-A systems with a conical intersection. Non-adiabatic transfer yields for various same-symmetry accidental conical intersection topographies are also observed to be significantly large in \cite{farfan2020systematic}.

\begin{acknowledgments}
This work is supported by grant ANR-21-CE30-0060-01 (COCOTRAMOS project) from Agence Nationale de la Recherche. LGM acknowledges support from grants 2021/04107-0, São Paulo Research Foundation (FAPESP), CNPq (305257/2022-6), and the US Air Force Office of Scientific Research (Grant FA9550-23-1-0666).
\end{acknowledgments}

\appendix
\section{\label{ssec:basis}Calculation details}

The present MRCI-SD calculations are performed using the ($2s$, $2p$) Rb atomic Gaussian basis set provided with the ECP36SDF effective core potential. The Rb basis set is extended with the set of diffuse Gaussian orbitals (with exponents in parenthesis) $1s (0.007181)$, $1p (0.004458)$, $2d (0.09,0.06)$, $1f (0.09)$, which are optimized to reproduce the depth of the Rb$_2$ electronic ground state potential. The CPP involves the Rb$^+$ static dipole polarizability (9.245 a.u.), and a cut-off radius $\rho= 0.225808a_0$, optimized for ionization potential (IP) of Rb (33690.81 cm$^{-1}$) \cite{NIST-ASD}. In Table \ref{tab:supRb2n}, we list the computed equilibrium distance $R_e$ and well-depth $D_e$ of the lowest gerade and ungerade states of Rb$_2$ and Rb$_2^+$. 

\setlength{\tabcolsep}{15.0pt}
\begin{table}[h]
\centering
\begin{tabular}{lcc}
\hline
 &  &  \\
State & $R_{e}$(a.u.) & $D_{e}$(cm$^{-1}$)  \\
 &  &  \\
\hline
 &  &  \\
Rb$_{2}(X^{1}\Sigma_{g}^{+})$ & 7.921 & 3993.98 \\
Rb$_{2}(X^{1}\Sigma_{g}^{+})$,  \cite{strauss2010}  & 7.956 & 3993.59  \\
Rb$_{2}(a^{3}\Sigma_{u}^{+})$ & 11.406 & 248.01  \\
Rb$_{2}(a^{3}\Sigma_{u}^{+})$, \cite{strauss2010} & 11.515  & 241.50  \\
 &  &  \\
Rb$_2^+$$(^{2}\Sigma_{g}^{+})$ & 9.12 & 6121.51 \\
Rb$_2^+$$(^{2}\Sigma_{u}^{+})$ & 22.82 & 80.26 \\
\hline
\end{tabular}
\caption{\label{tab:supRb2n}
Equilibrium distance, $R_{e}$, and well depth, $D_{e}$ of the Rb$_2$ and Rb$_2^+$ PECs computed in this work, and compared with a recent experimental result \cite{strauss2010}. }
\end{table}

\setlength{\tabcolsep}{12.0pt}
\begin{table}[h]
\centering
\begin{tabular}{llll}
\hline
 &  &  & \\
State & $\theta$ (\textdegree)& $R_e$ ($a_0$) & $D_{e}$(cm$^{-1}$)   \\
 &  & & \\
\hline
 &  & & \\
Rb$_3^+$(1$^1$A$_1$) & 60 & 8.77 & 13522 \\
Rb$_3^+$(1$^1$A$_1$) \cite{smialkowski2020} & 60 & 8.85 & 13258 \\
Rb$_3^+$(1$^1$B$_2$) & 180 & 11.76 & 1414 \\
Rb$_3^+$(2$^1$A$_1$) & 60  & 15.71 & 367 \\
Rb$_3^+$(1$^3$B$_2$) & 180 & 9.29 & 9444 \\
Rb$_3^+$(1$^3$B$_2$) \cite{smialkowski2020} & 180 & 9.93 & 9342 \\
Rb$_3^+$(1$^3$A$_1$) & 60 & 9.89 & 6494 \\
Rb$_3^+$(2$^3$B$_2$) & 60  & 22.71 & 164 \\
\hline
\end{tabular}
\caption{\label{tab:supRb3p}
Angle $\theta$, distance $R_e = R_{12} = R_{23}$, and well depth, $D_e$, of the minimal energy configuration for the six ITG states of Rb$_3^+$, correlated to the Rb$^+$$+$Rb($5s$)$+$Rb($5s$) limit taken as the origin of energy. Another theoretical value is available for 1$^1$A$_1$ and 1$^3$B$_2$ \cite{smialkowski2020}. }
\end{table}

The minimal energy configuration for the six ITG Rb$_3^+$ states, correlated to the Rb$^+$+Rb($5s$)+Rb($5s$) limit are reported in Table \ref{tab:supRb3p}. The origin of energies, taken at the Rb$^+$$+$Rb($5s$)$+$Rb($5s$) limit, is determined from the IP of Rb \cite{NIST-ASD}. The value of Rb dipole polarizability, $\alpha_d=322.1$~a.u., reported in the manuscript, is obtained from the fits on the Rb$_2^+$, $^2\Sigma_g^+$ and $^2\Sigma_u^+$ PECs in the long-range, from 100--500$a_0$. These fits use IP of Rb and quadrupole polarizability $\alpha_q$ = 6495 a.u. (= $2C_6$), taken from \cite{cote2016ultracold}.

\section{\label{ssec:singletPES}Potential energy surface of the 2$^1A$\textquotesingle~and 3$^1A$\textquotesingle~states}

\begin{figure}[h!]
\begin{tabular}{c}
\includegraphics[scale=0.275]{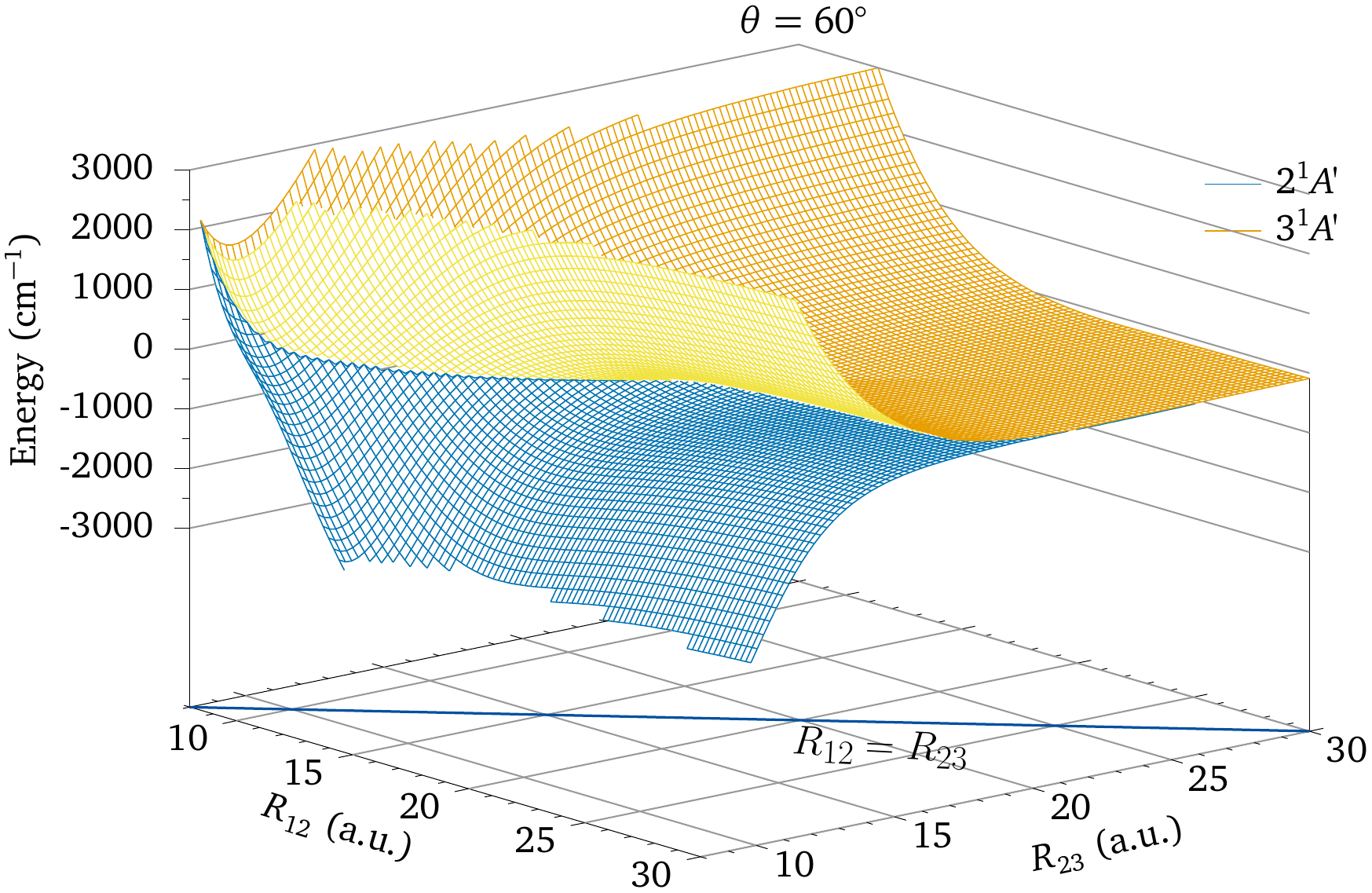}
\end{tabular}
\caption{PESs of the 2$^1A$\textquotesingle~and 3$^1A$\textquotesingle~Rb$_3^+$ states at $\theta=60$\textdegree~, and for $R_{12}$ and $R_{23}$ in the range $8a_0-30a_0$. The degeneracy of the PESs occurs for $R_{12} = R_{23}$ in the ITG, and is marked by the diagonal in the horizontal plane. }
\label{fig:singlet_PES}
\end{figure}

Figure \ref{fig:singlet_PES} displays the PESs of 2$^1A$\textquotesingle~and 3$^1A$\textquotesingle~Rb$_3^+$ states, thus completing Fig. \ref{fig:PESd1} for the triplet states displayed in the main text. The degeneracy of the PESs at $\theta=60$\textdegree~ and for all distances $R = R_{12} = R_{23}$ is visible, thus inducing JT coupling.

\section{\label{ssec:adtg30fr}Rb$_3^+$ PECs in the D$_>$ region: ADTG states for $R^c_{12}=30.00 a_0$}

\begin{figure}[ht]
\begin{tabular}{c}
\includegraphics[scale=0.535]{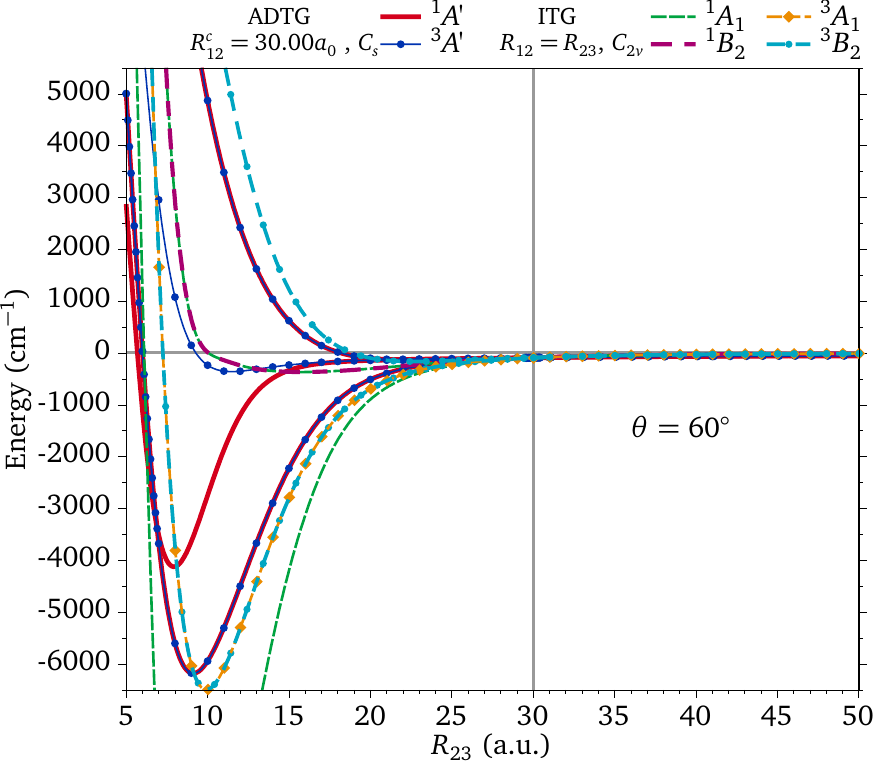}
\end{tabular}
\caption{For the larger $R_{23}$ range of Fig. \ref{fig:PESp1}, showing the ADTG for $R^c_{12}=30.00 a_0$ and ITG states at $\theta=60$\textdegree. }
\label{fig:lradtg30}
\end{figure}

As an exemplary case for D$_>$ region, ADTG for $R^c_{12}=30.00 a_0$ and ITG states at $\theta=60$\textdegree~are shown in Fig. \ref{fig:lradtg30}. It shows the curves for the larger range of Fig. \ref{fig:PESp1} of the manuscript. For $R^c_{12}=30.00 a_0$, Rb$_3^+$ singlet and triplet ADTG states in the short range, $R_{23} < R^c_{12}$, emulate the gerade and ungerade Rb$_2$ and Rb$_2^+$ PECs, see Fig. \ref{fig:Rb2np}. In the $R_{23} < R^c_{12}$ region, singlet and triplet curves correlated with each ADTG asymptote, $l_1$(Rb$_2^+$($^2\Sigma_g^+$) $+$ Rb) and $l_4$ (Rb$_2^+$($^2\Sigma_u^+$) $+$ Rb) become almost degenerate. As for the cases discussed for D$_<$ region, ADTG asymptotes, $l_2$ and $l_3$, associated with the neutral Rb$_2$ gerade and ungerade, are associated with one singlet and one triplet ADTG state, respectively.

\section{\label{ssec:nacmeJT}Non-adiabatic couplings in Rb$_3^+$}

\begin{figure}[ht]
\begin{tabular}{c}
\includegraphics[scale=0.645]{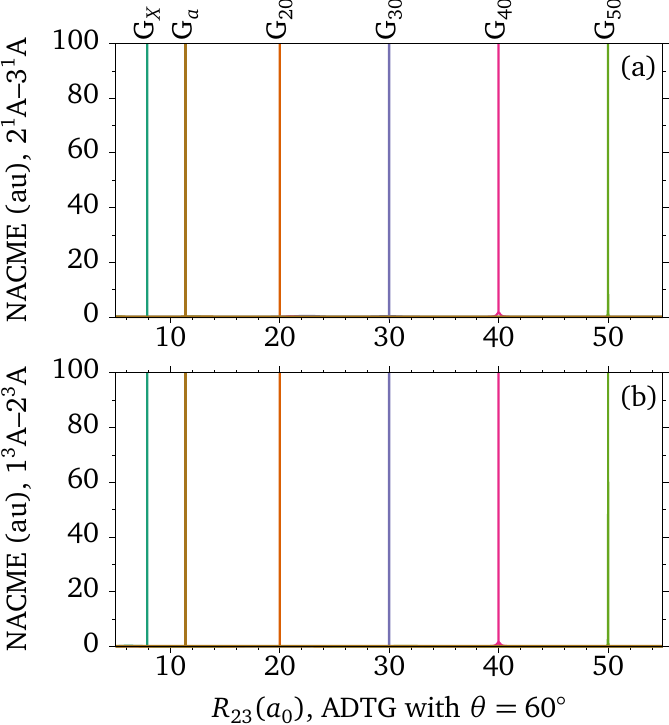}
\end{tabular}
\caption{(a) NACME as functions of $R_{23}$ for $\theta=60$\textdegree~and given values of $R^c_{12}$, (a) for JT-coupled singlet $2^1A'$--$3^1A'$ states and (b) for JT-coupled triplet $1^3A'$--$2^3A'$ states. The G labels refer to the peaks in the NACME: G$_X$ (resp. G$_a$), for $R^c_{12} = 7.90 a_0$ (resp. $R^c_{12} = 11.40 a_0$) equal to the equilibrium distance of $X^1\Sigma_g^+$ (resp. $a^3\Sigma_u^+ $) in Rb$_2$. The others correspond to some arbitrary values of  $R^c_{12}$ (= 20, \ldots, 50$a_0$). }
\label{fig:nacme}
\end{figure}

Non-adiabatic coupling matrix elements (NACME) is computed by the finite difference method through the MOLPRO routine DDR \cite{werner1981mcscf,MOLPRO-WIREs}. For a given trimer geometry, the procedure uses electronic wavefunctions computed at slightly displaced geometries to evaluate their derivatives. For the displacement increment $\Delta R = 0.0001 a_0$, absolute values of NACME for the JT-coupled singlet and triplet states are displayed in Fig. \ref{fig:nacme} (a) and (b) for fixed distances $R_{12}^c$ and $\theta=60$\textdegree, and depending on $R_{23}$. The labels G$_X$ and G$_a$ correspond to equilateral triangle geometries with internuclear distances equal to the equilibrium distance $R_e$ of $X^1\Sigma_g^+$ and $a^3\Sigma_u^+$Rb$_2$ states, and the other labels G$_{R_{12}^c}$ for arbitrary values of $R_{12}^c$. They locate the place where NACME becomes arbitrarily large ($> 100$ au), illustrating the presence of conical intersections and strong non-adiabatic couplings between states. We checked that for large internuclear distances ($\sim$100$a_0$), NACMEs remain strong at equilateral triangle geometries. We also verified that NACME with the singlet or triplet states that do not participate in the JT coupling are small ($< 0.50$ au) at any energetically accessible geometry. Accidental crossing between the first and second singlet states of Rb$_3^+$ occurs in the repulsive part of the trimer potential, making it insignificant. However, it could be responsible for the internal conversion in the other cases where crossing takes place in the attractive potential regions. Intersystem crossing between charged trimer singlet and triplet states, on the other hand, would happen only in the presence of strong spin-orbit couplings.

\section{\label{ssec:gennak}Generalization to the other alkali homo-nuclear systems}

\begin{figure*}[t]
\begin{tabular}{c}
\includegraphics[scale=0.65]{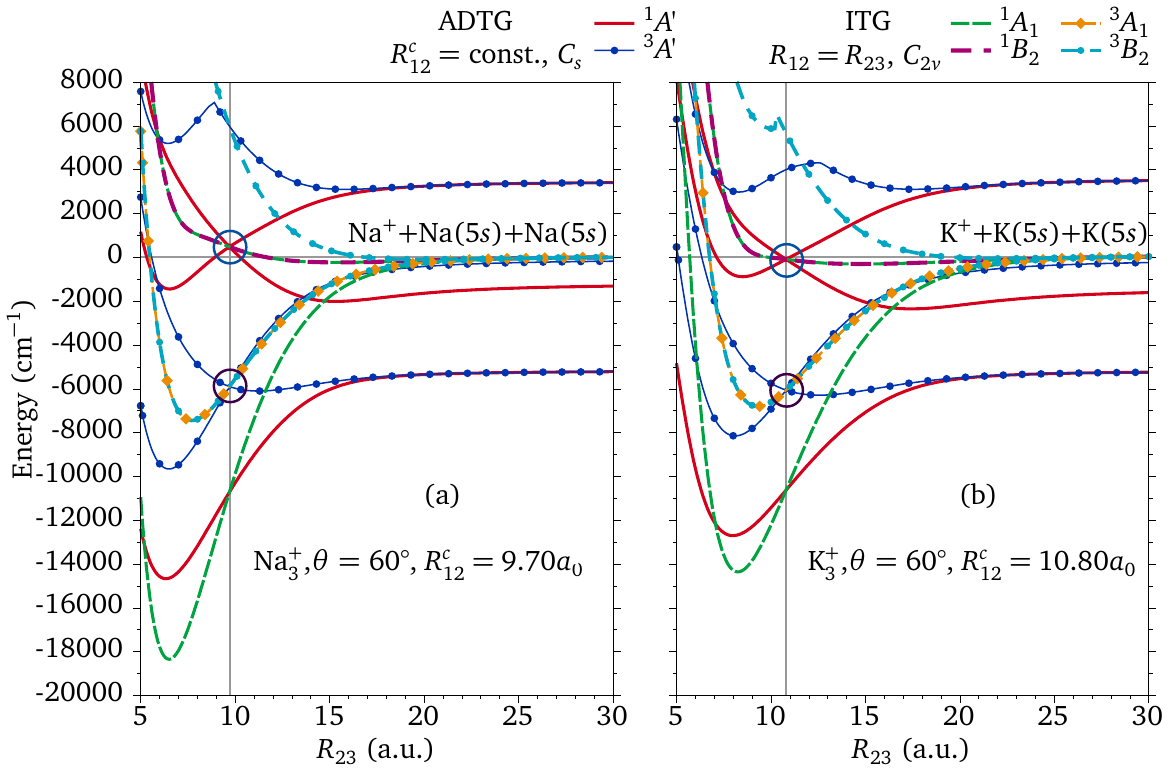}
\end{tabular}
\caption{(a) Na$_3^+$ ADTG and ITG PECs for $\theta$=60\textdegree. ADTG states are shown for $R_{12}^c = 9.70a_0$ ($R_e$ of Na$_2$$^3\Sigma_u^+$). (b) Same as (a) for K$_3^+$ with $R_{12}^c = 10.80a_0$ ($R_e$ of K$_2$$^3\Sigma_u^+$).}
\label{fig:appa}
\end{figure*}

Fig. \ref{fig:appa} shows Na$_3^+$ (a) and K$_3^+$ (b) ADTG and ITG PECs at $\theta$=60\textdegree~to demonstrate the occurrence of JT coupling as a general phenomenon in the singly-charged homo-nuclear alkali tri-atomic systems. These calculations are also performed using the MOLPRO package at the MRCI-SD level of theory. One-electron ECP+CPP atomic basis functions, ECP10SDF for Na, and ECP18SDF for K, and their optimized parameters are taken from \cite{shammout2023modeling}.

\end{document}